# Understanding Social Support Needs in Questions: A Hybrid Approach Integrating Semi-Supervised Learning and LLM-based Data Augmentation

**Junwei Kuang**, Ph.D.
Beijing Institute of Technology
Beijing, China, 100081
Email: kuangjw@bit.edu.cn

**Liang Yang**, Ph.D.
Beijing Institute of Technology
Beijing, China, 100081
Email: yl0119@outlook.com

**Shaoze Cui**, Ph.D.
Beijing Institute of Technology
Beijing, China, 100081
Email: shaoze-cui@foxmail.com

**Weiguo Fan**, Ph.D.
The University of Iowa
Iowa, USA
Email: weiguo-fan@uiowa.edu

Please send comments to Junwei Kuang at kuangjw@bit.edu.cn

# Understanding Social Support Needs in Questions: A Hybrid Approach Integrating Semi-Supervised Learning and LLM-based Data Augmentation

## Abstract

Patients are increasingly turning to online health Q&A communities for social support to improve their well-being. However, when this support received does not align with their specific needs, it may prove ineffective or even detrimental. This necessitates a model capable of identifying the social support needs in questions. However, training such a model is challenging due to the scarcity and class imbalance issues of labeled data. To overcome these challenges, we follow the computational design science paradigm to develop a novel framework, Hybrid Approach for SOcial Support need classification (HA-SOS). HA-SOS integrates an answer-enhanced semi-supervised learning approach, a text data augmentation technique leveraging large language models (LLMs) with reliability- and diversity-aware sample selection mechanism, and a unified training process to automatically label social support needs in questions. Extensive empirical evaluations demonstrate that HA-SOS significantly outperforms existing question classification models and alternative semi-supervised learning approaches. This research contributes to the literature on social support, question classification, semi-supervised learning, and text data augmentation. In practice, our HA-SOS framework facilitates online Q&A platform managers and answerers to better understand users' social support needs, enabling them to provide timely, personalized answers and interventions.

# Understanding Social Support Needs in Questions: A Hybrid Approach Integrating Semi-Supervised Learning and LLM-based Data Augmentation

## 1 INTRODUCTION

Social support refers to the diverse forms of assistance, care, and resources that individuals receive from their social networks, including family, friends, coworkers, and community members (Vaux, 1988). Social support has positive impacts on health, such as improving depression treatment (Zepeda & Sinha, 2016), increasing life satisfaction (Trepte et al., 2015), and reducing internet addiction severity (Lin et al., 2018). With the proliferation of the Internet, there has been a notable shift towards online health communities (OHCs) such as PatientsLikeMe[1], WebMD[2], and HaoDoctor[3] for obtaining online social support (Huang et al. 2019; Zhou et al. 2023a). Approximately 73% of individuals seek health information online (Shandwick, 2018). On Facebook, more than 1.8 billion users participate in health-related discussions monthly, forming over 10 million communities (Martin, 2022). PatientsLikeMe attracts over 100,000 visitors per month[4], while nearly one-third of the online U.S. population turns to WebMD for answers to their health-related questions[5].

Social support has been extensively studied within the Information Systems (IS) community. Most research focuses on understanding the motivation behind why users provide social support. Commonly considered factors include reciprocity (Yan, 2020), the pursuit of social capital (Huang et al., 2019), and the desire to expand virtual relationships (Leimeister et al., 2008). Another stream focuses on exploring the benefits of receiving social support, including improvements in patients' mental health (Yan & Tan, 2014), the reduction of rural-urban health disparities (Goh et al., 2016), and enhanced medication adherence (Lekwijit et al., 2024).

---

[1] https://www.patientslikeme.com/
[2] http://www.webmd.com/
[3] http://www.haodf.com/
[4] https://www.similarweb.com/website/patientslikeme.com/#overview
[5] https://www.prnewswire.com/news-releases/have-a-health-related-question-webmd-will-provide-the-answer----just-ask-alexa-300418914.html

However, the social support that patients receive is not always beneficial and can, in some cases, be harmful. The effectiveness of such supports largely depends on its alignment with the asker's specific needs (Yan, 2018). Generally, social support needs can be classified into five categories (Chiu et al., 2015): emotional needs, informational needs, network needs, esteem needs, and tangible needs, as summarized in Table 1. When social support are misaligned, it may lead to marital discord (Brock & Lawrence, 2009), unsuccessful weight loss (Yan, 2018), and increased stress among workers (Gray et al., 2020). Consequently, it is critical to identify the social support needs in askers' questions. Such identification not only enables answerers to provide more relevant support but also facilitates the implementation of automated interventions, including answer retrieval (Liu et al. 2020a), expert recommendation (Kundu et al., 2020), and AI chatbot responses (You et al., 2023). Therefore, our research aims to propose and evaluate an automated approach to identify social support needs in questions.

| Table 1. Categories of social support needs | |
|---|---|
| **Category** | **Description** |
| Emotional needs | seeking comfort, encouragement, and psychological support. |
| Informational needs | seeking guidance or advice to solve problems. |
| Network needs | seeking connections with others or find groups with similar interests. |
| Esteem needs | seeking recognition of one's abilities from others. |
| Tangible needs | seeking concrete resources to address problems. |

Machine learning techniques offer a promising solution for identifying social support needs in health-related questions (Padmanabhan et al., 2022). However, these techniques typically require extensive and well-balanced labeled datasets for model training (Merdan et al., 2021; Simester et al., 2020). A significant challenge lies in the manual labeling process, which is not only labor-intensive but also demands specialized expertise to accurately categorize social support needs (Park & Ho, 2021). Furthermore, in online Q&A communities, the distribution of social support needs is highly imbalanced (Alasmari et al., 2023). While most users seek informational or emotional support, other critical types of support, such as network support, are equally important but significantly underrepresented in available data. This inherent data imbalance poses a major obstacle to the accurate identification of social support needs.

Semi-supervised learning (SSL) can address the challenge of data scarcity by leveraging a few labeled samples and a large amount of unlabeled data (Ebrahimi et al., 2020; Yu et al., 2024). A widely adopted SSL approach involves two key steps: first, training a model on the available labeled data; second, this model is utilized to generate pseudo-labels for the unlabeled data. The model is then further refined by training on a combined dataset of real and pseudo-labeled samples. However, in the context of question classification, existing SSL approaches have primarily relied on the question text alone to generate pseudo-labels, neglecting the potential of associated answer text to enhance the quality of these pseudo-labels (Duarte & Berton, 2023), where there is a strong semantic dependency between the topics of the original question and its subsequent answers (Zhang et al., 2024a). For example, an answer containing reassuring phrases such as "don't worry" suggests the presence of an emotional support need in the associated question. Therefore, we propose the first objective: *Incorporate answer text into an SSL framework to improve the quality of pseudo-label generation*.

To address the challenge of data imbalance, we employ text data augmentation techniques to generate additional data for the minority class of social support needs A prevalent method is back-translation, which entails translating the original text into a different language and then translating it back into the original language (Pellicer et al., 2023). However, existing text data augmentation methods often produce samples with limited diversity, thereby constraining their overall effectiveness (Bayer et al., 2022). The advent of large language models (LLMs), renowned for their exceptional performance in natural language generation tasks, has inspired researchers to explore their potential for text data augmentation (Thirunavukarasu et al., 2023). Despite their promise, one prominent issue is the hallucination problem inherent in LLMs, which may lead to inaccuracies in the class labels of generated samples (Chang et al., 2024). Such unreliable samples can negatively impact model training and performance. We propose the second objective: *Utilize LLMs to generate diverse and reliable samples for minority classes*.

Following the computational design science paradigm (Abbasi et al., 2024; Gregor & Hevner, 2013; Hevner et al., 2004), we propose a novel framework, Hybrid Approach for SOcial Support need classification (HA-SOS). HA-SOS leverages a small number of labeled samples, a large amount of unlabeled data derived through SSL, and LLM-generated data to enhance the performance of question classification models. There are two main novelties in HA-SOS's design. First, to improve the quality of pseudo-labels for unlabeled questions, HA-SOS incorporates an answer-enhanced semi-supervised learning approach. This approach includes a dynamic two-dimension (2D) interaction kernel that captures the typical sentence pair between question-and-answer texts, as well as a quality-aware attention mechanism that assigns greater weights to high-quality answers based on user feedback, such as "best answer" labels. Second, to ensure label reliability and sample diversity in the generated data, HA-SOS integrates an LLM-based text augmentation approach with a nearest neighbor-based sample evaluation method. This method selectively retains diverse and reliable generated samples for underrepresented classes.

Extensive experiments conducted on a real-world dataset from *Yahoo! Answers* demonstrate that HA-SOS outperforms various baseline question classification models, as well as alternative semi-supervised learning and text data augmentation approaches. In practice, our framework enables answerers to provide more tailored and relevant answers. Additionally, the platform can effectively identify social support needs, facilitating the efficient recommendation of answerers and improving the overall quality of interactions in online Q&A communities.

## 2 LITERATURE REVIEW

Our research is connected to the following four research streams. First, we review recent social support studies in IS to underscore the importance of identifying social support needs, and the computational design science paradigm to guide the development of new methods. Second, we review existing question classification models, as they share the core task with our study. Then, we review semi-supervised learning and text data augmentation, both of which are directly relevant to our objective of addressing challenges related to data scarcity and imbalance. Finally, we summarize existing research gaps and propose corresponding research questions.

### 2.1 Social Support Studies and Computational Design Science Guidelines

The Internet has significantly enhanced interpersonal connections, leading many individuals to increasingly seek online social support through OHCs as a complement to the limited offline support they receive from face-to-face interactions (Goh et al. 2016; Huang et al. 2019; Liu et al. 2020b; Zhou et al. 2022). As a prominent form of health information technology (HIT), OHCs have been extensively studied by IS scholars (Baird et al., 2020; J. Zhou et al., 2023a). To contextualize our study, we summarize studies related to social support in Table 2.

Existing studies predominantly employ empirical models to examine the antecedents and effects of social support-related behaviors. Regarding antecedents, factors such as social capital (Chen et al., 2019; Huang et al., 2019) and self-disclosure (Lee et al., 2023) significantly influence the provision of social support. In terms of effects, receiving social support has been shown to enhance patients' perceived empathy (Nambisan, 2011), member commitment (Yang et al., 2017), and weight-loss outcomes (Yan, 2020). Additionally, the impact of social support may vary depending on the patient's health condition (Yan & Tan, 2014).

| Table 2. Summary of Studies Related to Social Support | | |
|---|---|---|
| **Category** | **Reference** | **Finding** |
| Antecedents of providing social support | Leimeister et al. (2008) | Characteristics of virtual social relationships, such as the intensity of Internet use, significantly influenced the provision of social support. |
| | Huang et al. (2019) | Structural capital, relational capital, and cognitive capital positively impacted emotional support provision, while only cognitive capital positively affected informational support provision. |
| | Chen et al. (2019) | Structural social capital significantly enhanced the provision of both informational and emotional support. |
| | Chen et al. (2020) | Affective linguistic signals, including negative sentiment and linguistic style matching, positively influenced the provision of informational and emotional support. |
| Effects of receiving social support | Oh et al. (2014) | Companionship in social support positively impacted life satisfaction, while appraisal and esteem enhanced the sense of community. |
| | Yan and Tan (2014) | Both informational and emotional support significantly improved patients' health conditions. |
| | Yan (2020) | Received support directly improved weight-loss outcomes and indirectly influenced self-monitoring activities. |
| | Wang et al. (2020) | Emotional and informational support positively correlated with consumer involvement in online communities. |
| Predictive social support analytics | Zhou et al. (2023a) | Designed a compound hierarchical attention networks model to predict whether a post constitutes emotional support for the target. |
| | Lee et al. (2023) | Developed a BERT-based ensemble classifier to identify types of social support offered in users' comments. |
| | Alasmari et al. (2023) | The performance of three BERT models (BERT, RoBERTa, and DistilBERT) was compared in identifying three types of social support needs (informational, emotional, and social) in questions. |

Furthermore, a few research explores predictive analytics for social support provision. For instance, Zhou et al. (2023a) employed a deep learning model to distinguish between emotional support and auxiliary content in answers. Lee et al. (2023) designed an ensemble model to classify the types of social support provided in users' comments. These identified social support types are often utilized as variables in empirical studies (Yan, 2018; Yan & Tan, 2014).

However, the social support users receive online is not always beneficial and can occasionally be detrimental (Yan, 2018). Misaligned social support may lead to adverse outcomes, such as marital discord (Brock & Lawrence, 2009), unsuccessful weight loss (Yan, 2018), and increased stress among workers (Gray et al., 2020). These findings highlight the critical need to accurately identify the social support needs embedded in askers' questions. Despite the significant contributions, few studies focus on detecting social support needs within questions. Moreover, the performance of existing models for identifying social support needs has been limited due to challenges such as scarce data and class imbalance (Alasmari et al., 2023).

The design science paradigm provides a framework with prescriptive guidelines for designing, developing, and evaluating innovative IT artifacts tailored to address critical societal challenges (Rai, 2017). Following the computational design science paradigm, three guidelines are followed. First, the artifact's design can be informed by critical domain requirements or distinctive characteristics of the problem space, ensuring that the artifact is grounded in real-world needs and practical considerations. Second, the novelty and efficacy of the artifact must be rigorously demonstrated by evaluating its technical performance against state-of-the-art (SOTA) approaches using well-established quantitative metrics. Finally, the artifact should make a meaningful contribution to the IS knowledge base, providing insights and frameworks that can guide future research in related domains. In this study, health Q&A platforms exhibit distinctive features such as multiple answers per question and asker-designated best answers, motivating the development of advanced IT artifacts to identify social support needs in questions.

## 2.2 Question Classification

Question classification (QC) refers to the process of categorizing questions according to the type of answer they expect (Mirzaei et al., 2023). QC has broad applications across diverse domains, such as identifying product requirements in customer inquiries (Rana et al., 2023), and discerning whether the intent behind a question is positive, negative, or neutral (Mirzaei et al., 2023). This categorization results can helps narrow down potential answer candidates (Silva et al., 2011) and facilitates the generation of automated answers (Anhar et al., 2019). Existing QC methods can be broadly classified into three categories, as illustrated in Table 3: rule-based methods, machine learning-based methods, and deep learning-based methods.

| Table 3. Summary of Studies Related to Question Classification | | |
|---|---|---|
| **Category** | **Reference** | **Finding** |
| Rule-based | Ray et al. (2010) | Developed a semantic method using WordNet and Wikipedia to identify whether a question involves location, numeric values, entities, etc. |
| | Madabushi and Lee (2016) | Utilized syntactic analysis and named entity recognition to detect entities like locations and numeric values in questions. |
| Machine learning | Zhang and Lee (2003) | Proposed an SVM model with a tree kernel to identify whether a question includes location or numeric references. |
| | Chen et al. (2012) | Applied semi-supervised learning to classify user question intentions into objective, subjective, and social categories. |
| | Mohasseb et al. (2018) | Developed a grammar-based approach for feature extraction, employing four machine learning methods to identify six question intent types (e.g., confirmation, factoid). |
| Deep learning | Liu et al. (2019) | Built an attention-based deep learning model for classifying the intent of Chinese questions, including comparison and enumeration. |
| | Faris et al. (2022) | Designed a deep learning model to classify questions within medical specialties (e.g., diabetes, child health). |
| | Yilmaz and Toklu (2020) | Trained a word embedding model using Turkish corpus and combined with deep learning to classify questions. |

Rule-based methods rely on predefined rule systems to classify questions, often utilizing manually crafted rules or natural language processing (NLP) techniques (Madabushi & Lee, 2016; Ray et al., 2010). While effective in specific contexts, these methods require significant manual efforts to create and maintain rules, making them costly and limited in generalizability. Machine learning-based methods address these limitations by employing data-driven techniques to train models. Common approaches include support vector machines (SVM) (Zhang & Lee, 2003) and decision trees (DT) (Mohasseb et al., 2018). However, these methods still depend on manually constructed feature sets. Deep learning-based methods have recently gained prominence due to their ability to automatically extract features from raw data, eliminating the need for manual feature engineering. Popular techniques include convolutional neural networks (CNN) (Liu et al., 2019) and long short-term memory (LSTM) networks (Faris et al., 2022), which have demonstrated superior performance in developing question classification models.

The social support needs classification is essentially a branch of question classification. Existing techniques typically require large volumes of labeled data to achieve satisfactory performance (Faris et al., 2022; Mallikarjuna & Sivanesan, 2022). However, to identify social support needs, obtaining sufficient labeled data presents significant challenges. First, the subjective nature of social support, along with variability in user expressions, necessitates the involvement of professional data labelers, which makes the labeling process costly and labor-intensive. Second, social support needs are often distributed in a highly imbalanced manner, with many users' needs focusing on informational and emotional needs, while other categories, such as network needs, receive less attention. This imbalance, coupled with the limited availability of labeled data, hampers the model's performance. To mitigate these challenges of data scarcity and imbalance, semi-supervised learning and text data augmentation techniques have emerged as promising solutions, which will be reviewed in the following sections.

### 2.3 Semi-Supervised Learning

SSL seeks to enhance model performance by leveraging both a small amount of labeled data and a large amount of unlabeled data (Shin et al., 2013). This approach has been applied to a wide range of tasks, including emotion recognition (Kang et al., 2021), detection of drug trafficking activities (Hu et al., 2023), identification of metastatic prostate cancer (Merdan et al., 2021).

Existing SSL methods can be categorized into self-training (Emadi et al., 2021), graph-based (Dai et al., 2023), and generative model-based methods (Yang et al., 2023). Specifically, self-training operates by iteratively generating pseudo-labels for unlabeled data using the model's predictions and incorporating these pseudo-labeled samples into the training set (Emadi et al., 2021). Graph-based methods transform data samples into graph representations, utilizing label propagation techniques to annotate unlabeled data effectively (Wang et al., 2023). Finally, generative model-based methods assume that data originates from a specific latent distribution and use the estimated distribution to infer labels for unlabeled data (Yang et al., 2023).

Among these three SSL approaches, graph-based and generative model-based methods, while powerful, often demand substantial computational resources due to their reliance on complex calculations and extensive model training (Yang et al., 2023). In contrast, self-training is more flexible and computationally efficient. This study thus adopts self-training as the foundational framework for developing a new SSL approach. We summarize related studies on self-training in text classification in Table 4. These studies demonstrate how self-training improves model performance across various tasks (Ligthart et al., 2021).

However, previous studies have primarily focused on utilizing question texts while neglecting the potential of answer texts (Li et al., 2017). For instance, consider the question: "*I'm feeling very lonely after my diagnosis, what should I do?*" An answer such as "*You should try joining a local support group*" suggests a need for network support, whereas "*I know how hard this must be for you*" indicates a need for emotional support. While no studies have explicitly explored the use of answer texts to enhance question classification, prior research has demonstrated the effectiveness of leveraging associated comments to improve post-level classification (Wang et al., 2019; Wei et al., 2022). For example, Wei et al. (2022) utilized news comment data to improve the detection of false news. These findings motivate us to incorporate answer texts into the training process to enhance the quality of pseudo-labels for unlabeled questions.

| Table 4. Studies on Self-training Semi-supervised Learning in Text Classification | | | |
|---|---|---|---|
| **References** | **Self-training Labeling Approach** | **Predictive Task** | **Additional Information** |
| Pavlinek and Podgorelec (2017) | Using labeled documents to predict pseudo-labels for documents of unknown categories. | Document classification | No |
| Li et al. (2017) | Leveraging question text to predict pseudo-label for questions. | Question classification | No |
| Meng et al. (2018) | Iteratively bootstrapping the unlabeled data to obtain high-quality deep neural models. | Document classification | No |
| Dong and de Melo (2019) | Leveraging a multilingual model's own predictions on unlabeled non-English data to obtain additional information. | Cross-lingual text classification | No |
| Meng et al. (2019) | Pseudo-label data with high confidence was continuously added to the training. | Hierarchical text classification | No |
| Li et al. (2021) | Improving the prediction accuracy of pseudo labels by applying angular margin loss during self-training. | Document classification | No |
| Ligthart et al. (2021) | Using different ratios of labeled review to predict pseudo-label for unlabeled review. | Opinion spam classification | No |
| Cui et al. (2022) | Selecting high-confidence words from classification results and adding them to the training set. | Short text classification | No |
| Our method | Leveraging question text and answer text to predict pseudo-label for questions | Question classification | Yes |

Existing methods for analyzing question and answer data suffer from two limitations. First, they fail to explicitly capture key question-answer pairs. For example, if a user asks, "How can I manage anxiety during social events?" and the answer states, "Practicing deep breathing exercises before the event can help reduce anxiety," current approaches that concatenate or independently analyze texts miss this direct alignment. Second, existing approaches typically assign weights to answers based on their semantic relation to the question (Verma et al., 2021), overlooking the real user feedback. Given the answers labeled by the asker as the "best answer" truly reflect their social support needs, incorporating the "best answer" as prior knowledge into the model's attention calculation can help prioritize important answers to the asker's needs.

While semi-supervised learning effectively mitigates data scarcity by leveraging unlabeled samples, it does not inherently address data imbalance issues. This imbalance leads models to prioritize the classification accuracy of majority classes at the expense of minority classes. We explore data augmentation techniques to increase the representation of minority class samples.

## 2.4 Text Data Augmentation

To address this data imbalance issue, researchers often employ text data augmentation techniques, which generate additional training samples without the need for collecting new data (Bayer et al., 2022). Common text data augmentation methods include random swaps, paraphrasing, and back translation. The random swap method involves randomly exchanging the positions of words within a sentence to create new variations (Gao et al., 2019). Paraphrasing techniques, on the other hand, generate alternative versions of a sentence by replacing words with their synonyms or rephrasing the text (Mi et al., 2022). Back translation leverages translation tools to first translate a sentence into another language and then translate it back into the original language. This process exploits linguistic differences to generate new sentences with altered phrasing while retaining the original meaning (Bayer et al., 2022).

Despite their widespread use, these traditional augmentation methods often suffer from limited text diversity, which restricts their ability to generate sufficiently varied samples for training (Pellicer et al., 2023). To address this challenge, we propose leveraging advanced LLMs, which have demonstrated a remarkable ability to understand and generate textual representations that closely resemble human language (Thirunavukarasu et al., 2023), demonstrating strong performance in handling text-related tasks such as text data augmentation. For instance, Yoo et al. (2021) incorporated practical examples into the prompts of LLMs, enabling the generation of augmented data that closely resembles these examples. Chintagunta et al. (2021) employed GPT-3 to produce a substantial volume of medical dialogue data, subsequently training a medical dialogue summarization model. Sahu et al. (2022) leveraged GPT-3 to generate labeled training data, which significantly enhances the performance of intent classifiers.

However, existing LLM-based text data augmentation methods face two significant limitations. First, due to the phenomenon of hallucination, LLM-generated samples may deviate significantly from the true class labels. To mitigate this issue, researchers have proposed few-shot approaches to enhance the quality of generated samples (Kowshik et al., 2024). While these methods improve semantic coherence, the limited number of example samples in few-shot settings can lead to inconsistencies between the generated samples' labels and the actual labels. Second, LLMs may be constrained by the input examples during sample generation, leading to a lack of diversity or repetition in the generated samples. This may result in insufficient feature space during model training. To address this issue, Kowshik et al. (2024) enhance the diversity of generated samples by selecting highly differentiated examples as prompts for LLMs. Whitehouse et al. (2023) employed a multi-LLM ensemble approach to generate diverse data. While these methods improve sample diversity to some extent, they fail to consider the differences between generated samples and existing labeled samples, ultimately reducing overall sample diversity.

## 2.5 Research Gaps and Questions

The literature review identifies four research gaps. First, existing question classification studies focus on utilizing unlabeled question texts while neglecting the potential of answer texts. Second, current methods for combining question and answer texts typically involve simple concatenation, failing to capture the nuanced interaction patterns between questions and answers. Third, existing attention mechanisms fail to incorporate the fact that askers often select the answer that best meets their needs as the "best answer" from multiple answers, which provides valuable prior knowledge. Fourth, previous LLM-based text augmentation methods have largely overlooked the dual importance of reliability and diversity in the samples generated. To address these limitations, this study proposes the following research questions (RQs):

RQ1: How can we leverage the content of answers to improve the performance of a question-based model for social support needs classification?

RQ2: How can we capture the typical interaction patterns between question-and-answer texts to improve the accuracy and interpretability of question classification?

RQ3: How can we refine the attention weights assigned to each answer by incorporating real patient feedback (i.e., "best answer") to enhance question classification?

RQ4: How can we design a selection mechanism to ensure the reliability and diversity of samples generated by LLMs, thereby improving the effectiveness of model training?

## 3 PROBLEM FORMULATION

We consider a dataset obtained from an online Q&A community, which consists of a labeled dataset $\mathcal{D}_l$ with $n_l$ labeled samples and an unlabeled dataset $\mathcal{D}_u$ with $n_u$ unlabeled samples. Each sample $\mathcal{S}$ includes a question text $T^{(q)}$ and its associated $K$ answer texts $\{T^{(a),1}, \ldots, T^{(a),K}\}$. Among the multiple answers, one is typically marked by the asker as the "best answer." To account for this, we incorporate an indicator $b_k$ ($b_k = 0 \ or \ 1$) in the representation vector of the answer text, where $b_k = 1$ to denote the $k$-th answer is designated as the best.

Each question text $T^{(q)}$ contains of $m$ sentences $\{T_1^{(q)}, \ldots, T_m^{(q)}\}$, while each answer text $T^{(a)}$ contains $n$ sentences $\{T_1^{(a)}, \ldots, T_n^{(a)}\}$. The set of social support need types is $\mathcal{C}$, containing three categories $\{Informational, Emotional, Network\}$. Here, $Informational$, $Emotional$, and $Network$ represent the needs for informational support, emotional support, and network support, respectively. For instance, if a given question includes both informational and emotional support needs, its label $y$ is represented by the vector $[1, 1, 0]$.

To address this classification problem, we aim to develop a model $f_Q(\cdot)$ that accurately identifies the type(s) of social support needed within question. Given the challenges of data scarcity and class imbalance in this task, we propose a learning approach that combines semi-supervised learning with LLM-based text data augmentation. This approach leverages unlabeled data and generated samples to enhance the model's predictive performance.

Table 5 summarizes the important notations used throughout the paper.

**Table 5. Important Notations**

| Notation | Description | Notation | Description |
|---|---|---|---|
| $\mathcal{D}_l$ | The original labeled dataset with $n_l$ samples | $\mathcal{C}$ | The set of sample labels, including three types $\{Informational, Emotional, Network\}$ |
| $\mathcal{D}_u$ | The original unlabeled dataset with $n_u$ samples (without pseudo label) | $h_m^{(q)}$ | The vectorized representation of the $m$-th sentence in the question text |
| $\mathcal{D}_u^*$ | A filtered unlabeled dataset with reliable pseudo-labels | $h_n^{(a),k}$ | The vectorized representation of the $n$-th sentence in the $k$-th answer text |
| $\mathcal{D}_a$ | Datasets augmented by LLMs with $n_a$ samples | $w_k$ | The weight assigned to the $k$-th answer in the quality-aware attention layer |
| $\mathcal{D}_a^*$ | The dataset retained after evaluating the LLMs generated data, $\mathcal{D}_a^* \subseteq \mathcal{D}_a$ | $y_{c,l}$ | The true label value of the $l$-th sample in social support need category $c$ |
| $\mathcal{D}_f$ | The fusion dataset used to train the Q-based model, which contains the question texts in $\mathcal{D}_l$, $\mathcal{D}_a^*$, and $\mathcal{D}_u^*$ | $p_{c,l}$ | The predicted probability of the $l$-th labeled sample in social support need category $c$ |
| $\mathcal{S}$ | The sample in a dataset, which contains one question $T^{(q)}$ and $K$ answers $T^{(a),1}, \ldots, T^{(a),K}$ | $\tilde{p}_{c,l}$ | The predicted probability of the $l$-th unlabeled sample in social support need category $c$ |

## 4 HA-SOS FRAMEWORK DESIGN

Following the computational design science paradigm (Rai, 2017), we propose the Hybrid Approach for SOcial Support need classification (HA-SOS) framework, as shown in Figure 1. HA-SOS includes four main steps: (1) *Training Q&A-based model*, which takes the patient's questions and associated multiple answers as model input and the social support needs of the question as model output; (2) *Pseudo-label prediction*; (3) *LLM-based data augmentation*; and (4) *Training Q-based model*, which only takes the patient's questions as model input.

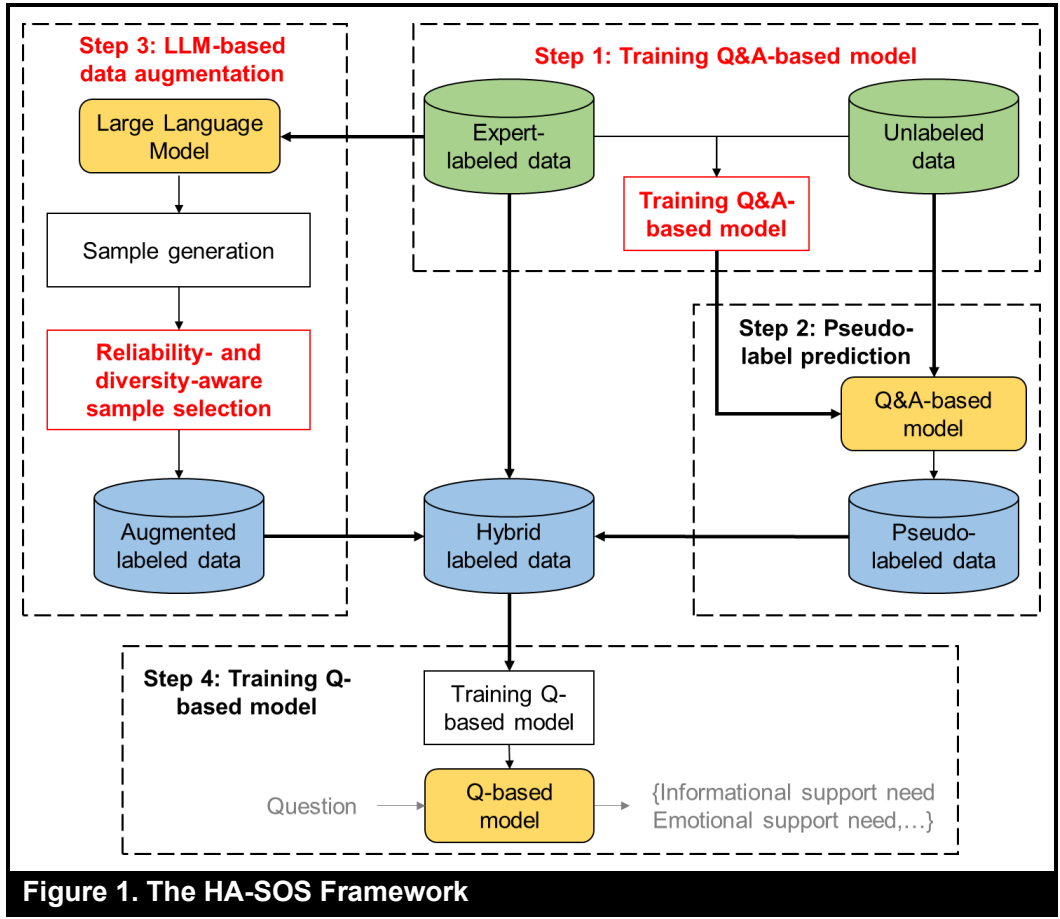

**Figure 1. The HA-SOS Framework**

HA-SOS begins by training a question classification model that utilizes both question and answer texts (Q&A-based) through a combination of expert-labeled and unlabeled data. Unlike previous semi-supervised question classification approaches, which rely solely on question texts (Q-based) to generate pseudo-labels for unlabeled questions. Once trained, the Q&A-based model generates pseudo-labels for unlabeled questions, effectively expanding the pool of labeled data. To address data imbalance issues, HA-SOS employs a reliability- and diversity-aware LLM-based data augmentation method to create additional high-quality samples for underrepresented classes. Finally, the pseudo-labeled samples, augmented data generated by LLMs, and original labeled samples are combined to train a refined Q-based question classification model that uses only question texts as input. We will detail each step next.

### 4.1 Training Q&A-based Model

To fully exploit the value of answer texts, HA-SOS introduces a new Q&A-based model, as shown in Figure 2. This model captures cross-position interactions between question-and-answer texts using a dynamic 2D kernel and employs a question-aware attention layer to assign varying weights to multiple answers associated with the question. The attention mechanism considers the user's perceived quality of each answer, such as the best answer selected for the question. Next, we provide a detailed, step-by-step explanation of each layer in the model.

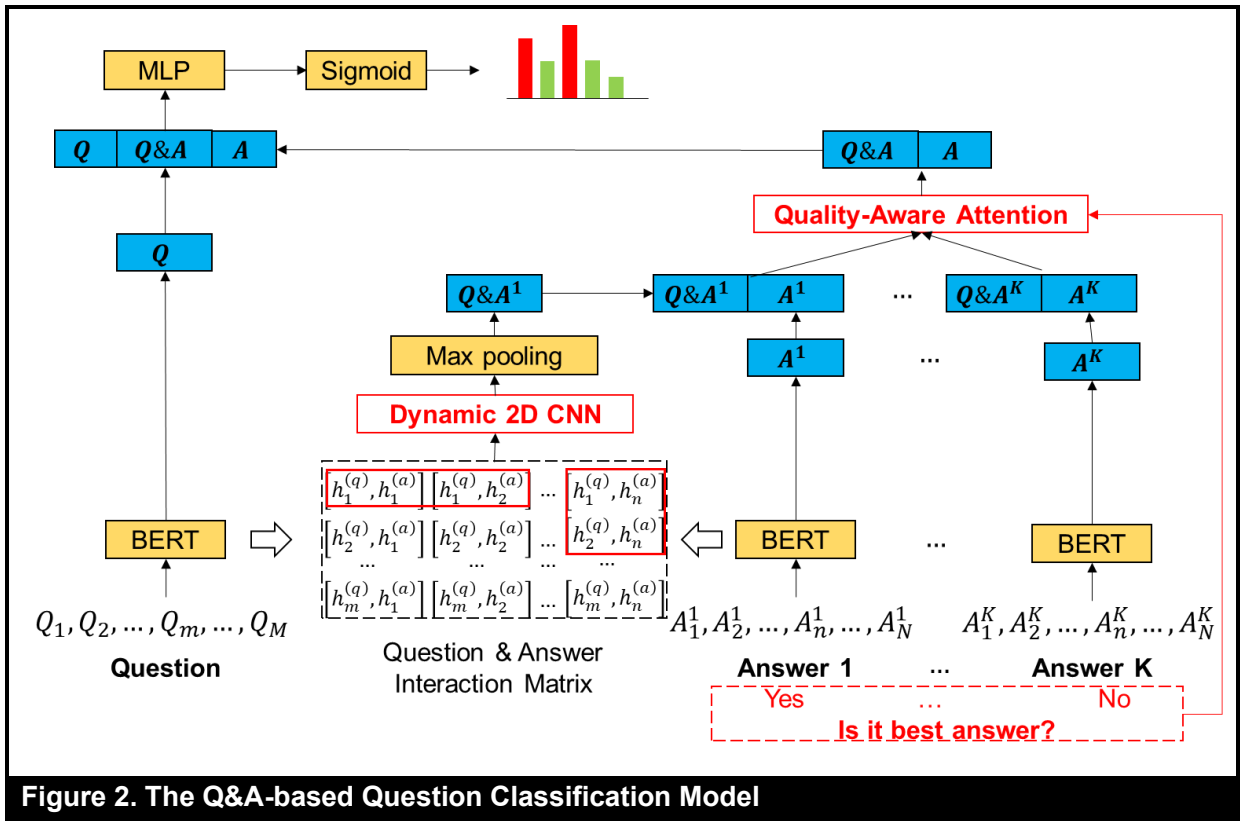

Figure 2. The Q&A-based Question Classification Model

#### 4.1.1 BERT Layer

We employ the Bidirectional Encoder Representations from Transformers (BERT) model (Devlin et al., 2019) to encode text in question-answering conversations. First, we encode the question text $T^{(q)}$ and answer text $T^{(a)}$ at the document level to obtain the global information of the text:

$$h^{(Q)} = \text{BERT}(T^{(q)}) \tag{1}$$

$$h^{(A),k} = \text{BERT}(T^{(a),k}) \tag{2}$$

where $T^{(q)}$ represents a question text, and $T^{(a),k}$ represents the $k$-th answer text corresponding to the question $T^{(q)}$. Second, while document-level encoding captures global context, it fails to capture fine-grained interactions. Furthermore, while word-level encoding can capture these nuanced details, it substantially increases computational complexity without delivering commensurate performance benefits. To balance these trade-offs, we perform text vectorization at the sentence level in conjunction with document-level encoding, thereby enabling a more nuanced understanding of the intricate interactions between question-and-answer texts:

$$\{h_1^{(q)}, h_2^{(q)}, \ldots, h_m^{(q)}\} = \text{BERT}\left(T_1^{(q)}, T_2^{(q)}, \ldots, T_m^{(q)}\right) \tag{3}$$

$$\{h_1^{(a),k}, h_2^{(a),k}, \ldots, h_n^{(a),k}\} = \text{BERT}\left(T_1^{(a),k}, T_2^{(a),k}, \ldots, T_n^{(a),k}\right) \tag{4}$$

where $m$ and $n$ represent the number of sentences in the question-and-answer texts, respectively. $h_m^{(q)}$ denotes the vectorized representation of the $m$-th sentence in the question text, and $h_n^{(a),k}$ represents the vectorized representation of the $n$-th sentence in the $k$-th answer text. If the number of sentences is fewer than $m$ or $n$, zero-padding is applied; if it exceeds $m$ or $n$, truncation is used to maintain consistent input size.

### 4.1.2 Q&A Interaction Layer

The Q&A interaction layer is designed to capture sentence-level interaction patterns between question-and-answer texts. Convolutional Neural Networks (CNNs) are highly effective for extracting such local patterns, as kernels of varying lengths can capture patterns of different granularities (Kim, 2014). For instance, in the sentence "I feel very depressed," a kernel length of 2 would extract word pairs such as "I feel," "feel very," and "very depressed."

However, traditional 2D convolution kernels are limited in their ability to extract cross-source word pairs—those that span both question and answer segments. To address this limitation, Zhu et al. (2021) introduced a 2D interaction kernel capable of processing two distinct data sources and capturing interactions between them. Despite its innovation, this approach imposes constraints: it requires that information from both sources occur simultaneously and be of equal length. In our context, the 2D interaction kernel is typically restricted to analyzing sentence pairs at the same positional index in the question-and-answer texts, as illustrated by the blue section in Figure 3.

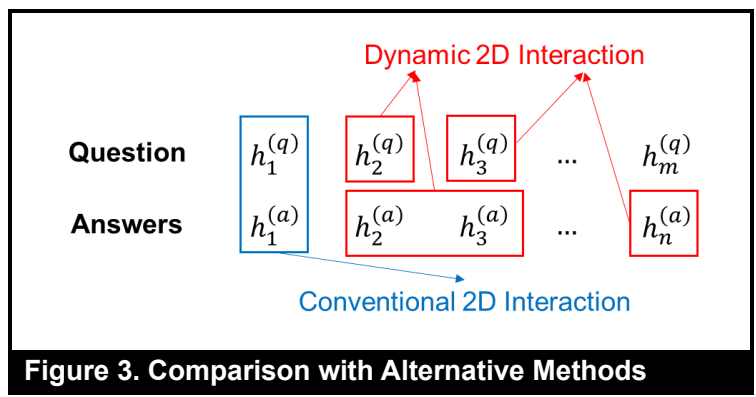

**Figure 3. Comparison with Alternative Methods**

In real-world online Q&A communities, sentences at different positions in question-and-answer texts often exhibit meaningful correspondences. For example, consider the question: "The weather is nice today, so I went out to meet some friends, but I feel very sad because I accidentally lost my phone. Can someone comfort me?" The corresponding answer is: "Don't be sad, maybe someone else picked it up and will contact you soon. Perhaps this is also a good opportunity to get a new phone." Here, the first sentence of the answer directly responds to the third sentence of the question, illustrating a cross-source and cross-position interaction pattern. To effectively capture such patterns, we propose a dynamic 2D interaction kernel. Figure 3 illustrates the differences between this kernel and alternative methods.

The process of extracting interactive information involves four steps. First, we construct an interaction matrix based on the embeddings of question-and-answer sentences. Second, we apply the dynamic 2D interaction kernel with varying sizes to capture diverse interaction patterns. Third, we use a maximum pooling strategy to extract the most significant activations for each Q&A interaction pattern. Finally, we concatenate these outputs to generate a comprehensive vectorized representation of the interactive content. The detailed calculation steps are as follows:

(1) Create an interaction matrix between sentences in the question and $k$-th answer:

$$M_{inter} = \begin{bmatrix} [h_1^{(q)}, h_1^{(a),k}] & \cdots & [h_1^{(q)}, h_n^{(a),k}] \\ \vdots & \ddots & \vdots \\ [h_m^{(q)}, h_1^{(a),k}] & \cdots & [h_m^{(q)}, h_n^{(a),k}] \end{bmatrix}_{m*n} \quad (5)$$

(2) Perform convolution operations using dynamic 2D interaction kernels:

$$M_{conv_{(i,j)}} = Conv_{(i,j)}(M_{inter}) \quad (6)$$

Kernels of varying sizes are employed to extract interaction patterns between sentences. For example, when $i = 2$ and $j = 2$, $Conv_{(2,2)}$ captures interaction relationships between two sentences in the question text and two sentences in the answer text.

(3) Capture the max activation of each kernel using a max pooling strategy:

$$M_{pooling_{(i,j)}} = Pooling\left(M_{conv_{(i,j)}}\right) \quad (7)$$

(4) Generate the vectorized representation of interactive content using concatenation:

$$h^{(q\&a),k} = Concat\left(Flatten\left(M_{pooling_{(i,j)}}\right)\right) \quad (8)$$

where $Concat(\cdot)$ denotes concatenation and $Flatten(\cdot)$ denotes the flattening operation.

### 4.1.3 Quality-aware Attention Layer

In our context, each question can receive multiple answers, necessitating an attention mechanism to assign varying weights to these answers based on their relevance. We compute these weights using the question-guided attention mechanism (Yang et al., 2016). The relevance score $s_k$ between the question representation $h^{(Q)}$ and the $k$-th answer representation $h^{(A),k}$ is computed using a single-layer MLP with a tanh activation function. This can be expressed as:

$$s_k = \tanh(h^{(Q)} W_s h^{(A),k} + b_s) \tag{9}$$

where $h^{(Q)}$ is the question vector, $h^{(A),k}$ is the $k$-th answer vector, $W_s$ is the weight matrix of the hidden layer, and $b_s$ is the bias term. This formulation captures the interaction between the question and each answer, producing a relevance score that reflects their semantic alignment.

The relevance scores $s_k$ are then normalized using the softmax function to obtain the attention weights $w_k$. This normalization ensures that the weights sum to one and can be interpreted as probabilities. The computation is defined as:

$$w_k = \frac{\exp(s_k)}{\sum_k \exp(s_k)} \tag{10}$$

These weights $w_k$ represent the relative importance of each answer with respect to the question. To enhance the efficacy of weight learning, we propose a quality-aware attention loss that integrates real feedback from the question asker to refine attention score. Specifically, if an answer is chosen as the best answer by the asker, its final weight $w_k$ should exceed that of other answers. The quality-aware attention loss is defined as:

$$\mathbb{1}(b_k = 1) \cdot \left( w_k - \max_{k=1,\ldots K} w_k \right)^2 \tag{11}$$

where $\mathbb{1}(b_k = 1)$ is an indicator function that equals 1 if the $k$-th answer is identified as the best answer and 0 otherwise, $w_k$ represents the weight assigned to the $k$-th answer, which is obtained by calculating the loss function, and $\max_{k=1,\ldots K} w_k$ denotes the maximum weight among all answers. This loss term penalizes the case where the weight of the best answer is not the highest weight.

The interaction information and the answer representations incorporate the weights learned from the quality-aware attention layer to aggregate the information of $K$ answers:

$$h^{(q\&a)} = w_1 \cdot h^{(q\&a),1} + \ldots + w_K \cdot h^{(q\&a),K} \tag{12}$$

$$h^{(A)} = w_1 \cdot h^{(A),1} + \ldots + w_K \cdot h^{(A),K} \tag{13}$$

### 4.1.4 Concatenation Layer

The layer consists of the representation of the question content $h^{(Q)}$, the representation of the interaction information $h^{(q\&a)}$, and the representation of the answer content $h^{(A)}$.

$$O^{Concat} = \text{Concat}(h^{(Q)}, h^{(q\&a)}, h^{(A)}) \tag{14}$$

This multi-dimensional representation enhances the model's ability to capture complex patterns and relationships, thereby improving its representational capacity.

### 4.1.5 MLP Layer

To achieve the final classification results, we employ an MLP layer:

$$O_{MLP}^{\mathcal{C}} = f_{MLP}^{\mathcal{C}}(W^{\mathcal{C}} * O^{Concat} + b^{\mathcal{C}}) \tag{15}$$

where $W^{\mathcal{C}}$ represents the weight coefficient to be learned in MLP, $b^{\mathcal{C}}$ represents the bias term, and $\mathcal{C}$ represents the type of label. To obtain the final predicted probabilities $p_{\mathcal{C}}$ for each label type, we transform the output $O_{MLP}^{\mathcal{C}}$ by applying the sigmoid function. This step ensures that the model's predictions are normalized to a range between 0 and 1, representing the likelihood of each support need being present in the question text. By leveraging this approach, the model is capable of simultaneously predicting multiple social support needs within a single question text.

### 4.1.6 Model training

We employ a self-training strategy from SSL to train the Q&A-based model. The process begins by training an initial model $f_{Q\&A}^0(\cdot)$ from labeled samples $D_l$. Next, $f_{Q\&A}^0(\cdot)$ is used to predict pseudo-labels for the unlabeled samples in $D_u$. These pseudo-labeled samples are then combined with the labeled samples in $D_l$ to train an updated Q&A-based model. This iterative process continues, where the improved model is used to predict pseudo-labels for unlabeled samples, and the model is retrained until convergence to the optimal model $f_{Q\&A}^*(\cdot)$.

The loss function $\mathcal{L}_{Total}$ for training the Q&A-based model as shown in Equation (16):

$$\mathcal{L}_{Total} = \lambda_L \mathcal{L}_{Label} + \lambda_U \mathcal{L}_{Unlabel} + \lambda_Q \mathcal{L}_{Quality\text{-}aware} \tag{16}$$

where $\lambda_L, \lambda_U, \lambda_Q$ represent the weights of the loss functions of different components. The loss function for these individual components is computed as follows:

The loss function for labeled samples is defined as:

$$\mathcal{L}_{Label} = -\frac{1}{n_l} \sum_{l=1}^{n_l} \sum_{c \in \mathcal{C}} y_{c,l} \log(p_{c,l}) \tag{17}$$

where $n_l$ represents the number of labeled samples, and $\mathcal{C}$ denotes the set of possible labels. The term $p_{c,l}$ corresponds to the model's predicted probability that the $l$-th question requires social support of type $c$, while $y_{c,l}$ represents the true label of the $l$-th question in one-hot encoding format. The one-hot encoding follows the structure $[Informational, Emotional, Network]$. For instance, a label $y_l = [1,1,0]$ indicates that the $l$-th question requires both informational and emotional support but does not require network support.

Following Sohn et al. (2020), the loss function for unlabeled samples is defined as:

$$\mathcal{L}_{Unlabel} = -\frac{1}{n_u} \sum_{l=1}^{n_u} \sum_{c \in \mathcal{C}} \mathbb{1}(\max(\tilde{p}_{c,l}, 1 - \tilde{p}_{c,l}) \geq \tau) \tilde{y}_{c,l} \log(\tilde{p}_{c,l}) \tag{18}$$

where $\tilde{p}_{c,l}$ represents the predicted probability of social support needs for the $l$-th unlabeled sample, and $\tilde{y}_{c,l}$ denotes the pseudo-label for the $c$-th category predicted by the model. The indicator function $\mathbb{1}(\cdot)$ equals 1 if $\max(\tilde{p}_{c,l}, 1 - \tilde{p}_{c,l}) \geq \tau$ and 0 otherwise, where $\tau$ is a hyperparameter. This ensures that only samples with confidence exceeding $\tau$ are utilized.

The quality-aware loss is developed to further refine the attention weights:

$$\mathcal{L}_{Quality\text{-}aware} = \frac{1}{n_l + n_u} \sum_{i=1}^{K} \sum_{l=1}^{n_l+n_u} \mathbb{I}(b_{i,l} = 1)\left(w_{i,l} - \max_{i=1,\ldots K} w_{i,l}\right)^2 \quad (19)$$

The core structure of Equation (19) remains consistent with Equation (11) and primarily serves to learn the weight distribution of different answers while ensuring that the best answer selected is assigned a highest weight. Since the overall loss is computed across both labeled and unlabeled datasets, the sample losses from these two parts are aggregated accordingly.

The overall training process of the Q&A-based model $f_{Q\&A}^*(\cdot)$ is summarized in Table 6.

| Table 6. Training Process of the Q&A-based Model | |
|---|---|
| **Input:** | Labeled dataset $D_l$; Unlabeled dataset $D_u$. |
| **Output:** | The optimal Q&A-based model $f_{Q\&A}^*(\cdot)$. |
| 1. | **# Supervised learning using labeled dataset $D_l$** |
| 2. | **while** the change in the combined loss function $\Delta(\mathcal{L}_{Label} + \mathcal{L}_{Quality-aware}) > \varepsilon$ **do:** |
| 3. |     **for** each sample $S = \{T^{(q)}|T^{(a),1}, \ldots, T^{(a),K}\}$ in $D_l$: |
| 4. |         Vectorize the question and associated answer texts using Equation (1) to (4); |
| 5. |         Obtain the vectorized representation of the Q&A interaction using Equation (5) to (8); |
| 6. |         Calculate the weights of different answers and generate vectorized representation according to Equation (9) to (13); |
| 7. |         Derive the concatenated vectorized representation using Equation (14); |
| 8. |         Use MLP to establish an output function for multi-label classification; |
| 9. |     **end for** |
| 10. |     Calculate the loss function using Equation (17) and (19) and update model parameters; |
| 11. | Obtain the initial Q&A-based model $f_{Q\&A}^0(\cdot)$ |
| 12. | **# Semi-supervised training using labeled dataset $D_l$ and unlabeled dataset $D_u$** |
| 13. | $D_u^* \leftarrow \emptyset, i = 0$; |
| 14. | **while** the performance change of the $f_{Q\&A}(\cdot)$ is greater than $\varepsilon$ **do:** |
| 15. |     Use the Q&A-based model $f_{Q\&A}^i(\cdot)$ to predict unlabeled samples in dataset $(D_u - D_u^*)$; |
| 16. |     Select samples that meet condition $\left(\max(\tilde{p}_{c,l}, 1 - \tilde{p}_{c,l}) \geq \tau\right)$ and add them to dataset $D_u^*$; |
| 17. |     **while** the total loss changes of $f_{Q\&A}^i(\cdot)$ is greater than $\varepsilon$ **do:** |
| 18. |         Train the $i$-th generation model $f_{Q\&A}^i(\cdot)$ using the synthetic dataset $D_l \cup D_u^*$; |
| 19. |     $i = i + 1$; |
| 20. | Obtain the optimal Q&A-based model $f_{Q\&A}^*(\cdot) \leftarrow f_{Q\&A}^i(\cdot)$. |

### 4.2 Pseudo-Label Prediction

Given the optimal Q&A-based model $f_{Q\&A}^*(\cdot)$, we generate pseudo-labels for unlabeled samples in $D_u$, thereby deriving a new dataset $D_u^*$ containing pseudo-labels. Specifically, for each unlabeled sample $S_u \in D_u$, the model predicts the probability of distribution over the possible labels $C$. The predicted probability for the $c$-th category is denoted as $\tilde{p}_{c,l}$, where $l$ indexes the sample. The pseudo-label $\tilde{p}_{c,l}$ is assigned based on a confidence threshold $\tau$:

$$\tilde{y}_{c,l} = \begin{cases} 1, \text{if } \max(\tilde{p}_{c,l}, 1 - \tilde{p}_{c,l}) \geq \tau \\ 0, \text{otherwise} \end{cases} \tag{20}$$

where $\tau$ is a predefined confidence threshold (e.g., $\tau = 0.9$). This ensures that only high-confidence predictions are retained as pseudo-labels. The resulting dataset $D_u^*$ consists of samples from $D_u$ with their corresponding pseudo-labels:

$$D_u^* = \{(S_u, \tilde{y}_{c,l}) | S_u \in D_u, \tilde{y}_{c,l} \text{ satisfies } \max(\tilde{p}_{c,l}, 1 - \tilde{p}_{c,l}) \geq \tau\} \tag{21}$$

The dataset $D_u^*$ is then combined with the labeled dataset $D_l$ to form a synthetic dataset $D_l \cup D_u^*$, which can help train the subsequent Q-based model $f_Q(\cdot)$.

### 4.3 LLM-based Data Augmentation

To mitigate this issue of class imbalance, we leverage LLMs to generate additional samples. Specifically, we design a tailored prompt, as detailed in Appendix A, to guide the LLMs to produce relevant and diverse data. A persistent challenge with LLMs is the phenomenon of "hallucination," where the model generates outputs that lack factual accuracy or grounding. Therefore, we evaluate and select LLM-generated samples with reliability and diversity.

Let the set of samples generated by the LLMs be denoted as $\mathcal{D}_a$. For each sample $S_i^a (i = 1,2, \ldots n_a)$ in $\mathcal{D}_a$, we compute its similarity to the real samples $S_j^l (j = 1,2, \ldots, n_l)$ in the labeled dataset $\mathcal{D}_l$, using cosine similarity $\cos\_\text{sim}(S_i^a, S_j^l)$. Based on this similarity, we select the top $k$ most similar samples from $\mathcal{D}_l$, forming a subset $\mathcal{D}_{l,a}^k = \{S_{i,1}^l, S_{i,2}^l, \ldots S_{i,k}^l\}$. We then evaluate the consistency and diversity of each generated sample $S_i^a$ using the following metrics:

$$Consistency(S_i^a) = \frac{\left|\{S_i^l \in \mathcal{D}_{l,a}^k : L_{S_i^l} = L_{S_i^a}\}\right|}{k} \quad (22)$$

$$Diversity(S_i^a) = 1 - \frac{1}{k} \sum_{S_i^l \in \mathcal{D}_{l,a}^k} \cos\_sim(S_i^a, S_i^l) \quad (23)$$

where $Consistency(\cdot)$ measures the degree of label consistency between the generated sample $S_i^a$ the labeled samples in $\mathcal{D}_{l,a}^k$. A higher consistency score indicates that $S_i^a$ aligns well with the ground-truth labels, reducing the risk of hallucination. Furthermore, $Diversity(\cdot)$ quantifies the dissimilarity between $S_i^a$ and the labeled samples in $\mathcal{D}_{l,a}^k$. Lower average similarity scores indicate greater diversity, which helps build a more robust and generalizable model.

To balance consistency and diversity in the LLM-generated samples, we introduce a comprehensive scoring mechanism to select suitable samples:

$$Score = \delta * Consistency(\cdot) + (1 - \delta) * Diversity(\cdot) \quad (24)$$

where $\delta$ is a weighting parameter that controls the trade-off between consistency and diversity. Based on this score, all LLM-generated samples are ranked, and only those exceeding a predefined threshold are selected to form the ultimate augmented dataset $\mathcal{D}_a^*$.

### 4.4 Training Q-based Model

The Q-based model in our framework is designed for multi-label classification. By leveraging pseudo-label predictions from the Q&A-based model and incorporating augmented data generated by the LLMs, we construct a comprehensive dataset $\mathcal{D}_f$. This dataset combines samples from the labeled dataset $\mathcal{D}_l$, the pseudo-labeled dataset $\mathcal{D}_u^*$, and the augmented dataset $\mathcal{D}_a^*$. The dataset $\mathcal{D}_f$ will serve as the input for training the Q-based model $f_Q(\cdot)$.

Q-based model processes input data in multiple stages. First, the input text is passed through a Bi-LSTM layer to extract contextual textual features. These features are then fed into a MLP layer, which maps the extracted features to corresponding class labels. Unlike the Q&A-based model, the Q-based model does not utilize answer data or unlabeled data. Therefore, Equations (19) and (20) are not required. Instead, the Q-based supervised learning model $f_Q^*(\cdot)$ is iteratively trained using only the cross-entropy loss, as defined in Equation (17). Once trained, this model can be directly applied to identify the social support needs in new questions.

## 5 EMPIRICAL ANALYSES

### 5.1 Data Collection and Preprocess

We utilized the CHQ-SocioEmo dataset, which comprises over 22,000 question-answer pairs from *Yahoo! Answers* (Alasmari et al., 2023). Trained human experts meticulously annotated approximately 1,500 questions based on the definitions of the five types of social support. Each question could carry multiple labels, indicating the presence of various support needs. We filtered the unlabeled dataset to include samples relevant to health, discarding any entries with missing best-answer label. Then, we randomly selected 10,000 unlabeled samples. Table 7 summarizes the descriptive statistics of the dataset, revealing a notable imbalance of different social support needs. Following Alasmari et al. (2023), we excluded samples related to tangible support and merged esteem support into the emotional support needs category.

| Table 7. Description of the Dataset | | |
|---|---|---|
| Category | Labeled Data | Unlabeled Data |
| # of Samples | 1500 | 10000 |
| Social Support Needs | | |
|     Informational support needs | 995 | - |
|     Emotional support needs | 624 | - |
|     Network support needs | 444 | - |
|     Esteem support needs | 207 | - |
|     Tangible support needs | 15 | - |
| # of Answers per Question | 6.09 | 6.54 |
| Average length of question | | |
|     Sentence | 7.66 | 3.06 |
|     Words | 125.87 | 42.90 |
| Average length of answer | | |
|     Sentence | 5.89 | 3.85 |
|     Words | 81.12 | 54.12 |

## 5.2 Experiment Design

We designed six distinct experiments to comprehensively evaluate the performance of the proposed method. The first experiment compared our approach with state-of-the-art models from related works, including machine learning techniques, deep learning models, and LLMs, to determine whether our method offered a competitive advantage. The second experiment conducted an ablation study to assess the contribution of individual components of our method, such as the dynamic 2D interaction kernel and text data augmentation, to overall performance. The third experiment focused on classification performance across different labels, allowing us to analyze how the model performs across various labels. In the fourth experiment, we analyzed the sensitivity of key parameters, including the number of unlabeled samples and the answer data, using multiple experimental setups. The fifth experiment explored the interpretability of the model, enabling users to better understand the rationale behind its classification decisions. Finally, the sixth experiment assessed the model's generalization ability using a new prediction task (i.e., emotion classification), demonstrating its potential for broader applications.

To ensure transparency and facilitate reproducibility, we summarized the key hyperparameter settings for the proposed HA-SOS method in Table 8. Detailed hyperparameters for all comparison methods (machine learning, deep learning, and semi-supervised approaches) are reported in Appendix B. Importantly, all hyperparameters, including those for our proposed method, were tuned via grid search to ensure fair and reliable comparisons.

To ensure a robust evaluation of our model, we employed the ten-fold cross-validation technique to partition the dataset into training and testing sets (Wang et al., 2021). To comprehensively assess our model's performance, we incorporated four evaluation metrics: Precision, Recall, F1, and AUC. Given the skewed distribution of class labels, we computed

Precision, Recall, and F1 using a micro-averaging method.

| Table 8. Key Hyperparameter Settings Involved in the HA-SOS Model | | | |
|---|---|---|---|
| **Hyperparameter** | | | **Value** |
| Semi-supervised learning | Dynamic 2D interaction kernel | Convolutional kernel | [1,1], [1,2], [2,1], [2,2] |
| | | AdaptiveAvgPool2d | [4,4] |
| | | Dropout | 0.4 |
| | # of unlabeled data | | 10000 |
| | # of answers | | 65492 |
| | # of answers for a question | | 5 |
| | Threshold $\tau$ for pseudo-label inclusion | | 0.7 |
| | The weight of each component in the total loss | | $\lambda_L, \lambda_U, \lambda_Q$ |
| Text data augmentation | LLMs | Version type | GPT-4o |
| | | Total number of generated samples | 19999 |
| | | # of network support needs | 14164 |
| | | # of emotion support needs | 6778 |
| | | # of information support needs | 2882 |
| | The weight coefficient $\delta$ used to balance consistency and diversity | | 0.4 |
| | Threshold of sample selection $\eta$ | | 0.2 |
| Model Training | Batch size | | 64 |
| | Optimizer | | Adam |
| | Learning rate | | 0.01 |
| | Schedular | | LinearLR |
| | Hidden layer size | | 256 |

Micro-averaging aggregates outcomes across all classes to compute an overall average, ensuring that each instance contributes equally to the final metric. This approach is suitable for imbalanced datasets (Tarekegn et al., 2021). The calculation for these metrics is as follows:

$$Micro\text{-}Precision = \frac{\sum_{c \in \mathcal{C}} TP_c}{\sum_{c \in \mathcal{C}} TP_c + FP_c} \qquad (25)$$

$$Micro\text{-}Recall = \frac{\sum_{c \in \mathcal{C}} TP_c}{\sum_{c \in \mathcal{C}} TP_c + FN_c} \qquad (26)$$

$$Micro\text{-}F1 = 2 \cdot \frac{Micro\ Precision \times Micro\ Recall}{Micro\ Precision + Micro\ Recall} \qquad (27)$$

$$Micro\text{-}AUC = \int_0^1 Micro\ TPR(Micro\ FPR) d(Micro\ FPR) \qquad (28)$$

$$\text{Micro-FPR} = \frac{\sum_{c \in C} TP_c}{\sum_{c \in C} FP_c + TN_c} \tag{29}$$

$$\text{Micro-FPR} = \frac{\sum_{c \in C} TP_c}{\sum_{c \in C} TP_c + FN_c} \tag{30}$$

where $TP_c$ (True Positive) represents the number of correctly predicted samples in category $c$, while $TN_c$ (True Negative) denotes the number of correctly predicted samples in all other categories excluding $c$. $FP_c$ (False Positive) refers to the number of samples from other categories that were incorrectly classified as category $c$, and $FN_c$ (False Negative) represents the number of samples that belong to category $c$ but are misclassified into other categories.

### 5.3 Evaluation of Predictive Power

#### 5.3.1 Comparison of Question Classification Models

By comparing the proposed HA-SOS model with the seven other machine learning models listed in Table 9, our model demonstrates significant improvements across four evaluation metrics: Precision, Recall, F1, and AUC. Notably, HA-SOS achieves a Recall of 72.50% (exceeding baselines by 12.91%–43.28%) and an AUC of 73.75% (improving by 13.76%–36.65%). These gains mainly stem from a deep learning-based model, which has stronger representation ability.

| Table 9. Comparison of Machine Learning Models | | | | | | | | |
|---|---|---|---|---|---|---|---|---|
| Model | Precision | Imp. | Recall | Imp. | F1 | Imp. | AUC | Imp. |
| DT (Bilal et al., 2016) | 52.78%*** | 35.45% | 53.75%*** | 34.88% | 53.21%*** | 35.09% | 55.60%*** | 32.64% |
| KNN (Zhang & Lee, 2003) | 55.87%*** | 27.96% | 54.42%*** | 33.22% | 55.08%*** | 30.50% | 58.19%*** | 26.74% |
| LR (Li et al., 2023) | 54.43%*** | 31.34% | 54.10%*** | 34.01% | 54.25%*** | 32.50% | 56.98%*** | 29.43% |
| MLP (Hamza et al., 2021) | 59.33%*** | 20.50% | 58.76%*** | 23.38% | 59.00%*** | 21.83% | 61.51%*** | 19.90% |
| NB (Momtazi, 2018) | 51.35%*** | 39.22% | 50.60%*** | 43.28% | 50.87%*** | 41.30% | 53.97%*** | 36.65% |
| RF (Mohasseb et al., 2018) | 62.70%*** | 14.02% | 57.59%*** | 25.89% | 59.98%*** | 19.84% | 63.56%*** | 16.03% |
| SVM (Zhang & Lee, 2003) | 63.80%*** | 12.05% | 59.61%*** | 21.62% | 61.60%*** | 16.69% | 64.83%*** | 13.76% |
| **HA-SOS** | **71.49%** | - | **72.50%** | - | **71.88%** | - | **73.75%** | - |

Note: Significance levels are denoted by * (0.05), ** (0.01) and *** (0.001) respectively. Decision Tree: DT. K-Nearest Neighbor: KNN. Multi-Layer Perceptron: MLP. Random Forest: RF. Support Vector Machine: SVM. Imp, Improvement

This experiment aims to evaluate the performance of HA-SOS against state-of-the-art deep learning models in question classification, as shown in Table 10. HA-SOS achieves a Recall of 72.50% and an AUC of 73.75%, outperforming the best baseline Alasmari et al. (2023) by 11.93% and 11.62%, respectively. This significant improvement stems from HA-SOS's unique design: (1) its Q&A-based pseudo-labeling leverages contextual information from both questions and answers; (2) its reliability- and diversity-aware data augmentation addresses class imbalance.

Table 10. Comparison of Deep Learning Models

| Model | Precision | Imp. | Recall | Imp. | F1 | Imp. | AUC | Imp. |
|---|---|---|---|---|---|---|---|---|
| LSTM (Faris et al., 2022) | 58.36%*** | 22.50% | 58.33*** | 24.29% | 58.25%*** | 23.40% | 59.76%*** | 23.41% |
| Bi-LSTM (Anhar et al., 2019) | 58.62%*** | 21.95% | 60.98*** | 18.89% | 59.68%*** | 20.44% | 60.42%*** | 22.06% |
| CNN (Liu et al., 2019) | 60.49%*** | 18.18% | 55.92*** | 29.65% | 58.00%*** | 23.93% | 61.00%*** | 20.90% |
| LSTM+ATT (Xia et al., 2018) | 58.00%*** | 23.26% | 57.83*** | 25.37% | 57.80%*** | 24.36% | 59.45%*** | 24.05% |
| HAN (Yang et al., 2016) | 59.31%*** | 20.54% | 57.10*** | 26.97% | 58.05%*** | 23.82% | 60.34%*** | 22.22% |
| Tan et al. (2022) | 60.91%*** | 17.37% | 55.99*** | 29.49% | 58.24%*** | 23.42% | 61.28%*** | 20.35% |
| Liang et al. (2021) | 58.67%*** | 21.85% | 57.93*** | 25.15% | 58.21%*** | 23.48% | 59.96%*** | 23.00% |
| Alasmari et al. (2023) | 60.92%*** | 17.35% | 60.57*** | 19.70% | 60.63%*** | 18.56% | 62.13%*** | 18.70% |
| **HA-SOS** | **71.49%** | - | **72.50%** | - | **71.88%** | - | **73.75%** | - |

Note: Significance levels are denoted by * (0.05), ** (0.01) and *** (0.001) respectively.

To validate the effectiveness of HA-SOS in leveraging unlabeled data, we compared it with other SSL methods. As shown in Table 11, HA-SOS achieves superior performance, with a Recall of 72.50% and an AUC of 73.75%, outperforming all baselines. Notably, HA-SOS surpasses the suboptimal method--Guo and Li (2022) by 16.67% in Recall and 3.19% in AUC, demonstrating its ability to generate more accurate pseudo-labels and effectively utilize unlabeled data. The results highlight the importance of combining advanced pseudo-labeling techniques with robust data augmentation in semi-supervised learning for question classification.

Table 11. Comparison against other Semi-supervised Learning Methods

| Category | Model | Precision | Imp. | Recall | Imp. | F1 | Imp. | AUC | Imp. |
|---|---|---|---|---|---|---|---|---|---|
| - | **HA-SOS** | **71.49%** | - | **72.50%** | - | **71.88%** | - | **73.75%** | - |
| Clustering | Chavoshinejad et al. (2023) | 60.62%*** | 17.93% | 61.99%*** | 16.95% | 61.15%*** | 17.55% | 63.22%*** | 16.66% |
| Self-training | Guo and Li (2022) | 74.17% | -3.61% | 62.14%*** | 16.67% | 67.39%*** | 6.66% | 71.47%** | 3.19% |
| Graph-based | Cui et al. (2022) | 64.65%*** | 10.58% | 54.66%*** | 32.64% | 59.08%*** | 21.67% | 64.05%*** | 15.14% |

Note: Significance levels are denoted by * (0.05), ** (0.01) and *** (0.001) respectively. Imp, Improvement.

To further validate the effectiveness of HA-SOS, we compared it against popular LLMs such as GPT-4, Llama-3, and Yi, applied directly to the social support needs classification task. As shown in Table 12, HA-SOS outperforms these LLMs in Precision, F1, and AUC. While GPT-4 and Llama-3 achieve high Recall (81.92% and 89.79%), their Precision is significantly lower, indicating a tendency to over-predict positive instances. This discrepancy may arise because LLMs, trained on vast and diverse corpora, prioritize capturing a wide range of potential matches at the cost of introducing false positives. In contrast, HA-SOS's tailored semi-supervised framework, which combines Q&A-based pseudo-labeling and LLM-augmented data, ensures a better balance between Recall and Precision. By leveraging domain-specific labeled and unlabeled data, HA-SOS minimizes false positives while maintaining high sensitivity.

Table 12. Comparison against LLMs

| Model | Precision | Imp. | Recall | Imp. | F1 | Imp. | AUC | Imp. |
|---|---|---|---|---|---|---|---|---|
| GPT 4o + zero shot | 54.28%*** | 31.71% | 81.92% | -11.50% | 65.29%*** | 10.09% | 62.28%*** | 18.42% |
| GPT 4o + few shot | 55.46%*** | 28.90% | 71.75%* | 1.05% | 62.56%*** | 14.90% | 62.03%*** | 18.89% |
| Llama 3 + zero shot | 52.94%*** | 35.04% | 89.79% | -19.26% | 66.61%*** | 7.91% | 58.22%*** | 26.67% |
| Llama 3 + few shot | 50.93%*** | 40.37% | 88.63% | -18.20% | 64.69%*** | 11.11% | 55.08%*** | 33.90% |
| Yi + zero shot | 50.00%*** | 42.98% | 58.47%*** | 24.00% | 53.90%*** | 33.36% | 52.37%*** | 40.82% |
| Yi + few shot | 52.92%*** | 35.09% | 61.02%*** | 18.81% | 56.68%*** | 26.82% | 55.56%*** | 32.74% |
| **HA-SOS(Ours)** | **71.49%** | - | **72.50%** | - | **71.88%** | - | **73.75%** | - |

Note: Significance levels are denoted by * (0.05), ** (0.01) and *** (0.001) respectively.

### 5.3.2 Ablation Studies

Ablation studies were conducted to evaluate the contribution of each component in HA-SOS, as shown in Table 13. The removal of any component leads to a decline in performance. The SSL module has the most significant impact. Its absence results in a nearly 15% drop across all metrics, underscoring its critical role in leveraging unlabeled data for improved classification. The answer part and quality-aware attention layer also play important roles, with their removal causing a 2.78%–7.92% decrease in Recall and a 3.13%–5.45% drop in AUC. This highlights the value of incorporating answer texts and attention mechanisms for capturing nuanced contextual information. Interestingly, while text data augmentation has a relatively modest impact (2% decline), the constraints of label consistency and sample diversity are more critical. Removing these constraints leads to a 7.34%–9.40% drop in performance, suggesting that while LLM-generated data enhances the model, it also introduces noise. This emphasizes the need for rigorous evaluation and filtering of augmented samples to ensure quality.

| Table 13. Ablations Studies | | | | | | | | |
|---|---|---|---|---|---|---|---|---|
| Model | Precision | Imp. | Recall | Imp. | F1 | Imp. | AUC | Imp. |
| **HA-SOS (Ours)** | **71.49%** | - | **72.50%** | - | **71.88%** | - | **73.75%** | - |
| **W/o Semi-supervised learning** | 60.57%*** | 18.03% | 62.97%*** | 15.13% | 61.62%*** | 16.65% | 62.59% | 17.83% |
| W/o Answer part | 69.11%*** | 3.44% | 70.54%** | 2.78% | 69.75%** | 3.05% | 71.51% | 3.13% |
| W/o Quality-aware attention layer | 68.23%*** | 4.78% | 67.18%*** | 7.92% | 67.64%*** | 6.27% | 69.94% | 5.45% |
| W/o Dynamic 2D interaction kernel | 70.26%** | 1.75% | 65.00%*** | 11.54% | 67.36%*** | 6.71% | 70.30%*** | 4.91% |
| **W/o Text data augmentation** | 70.84%* | 0.92% | 70.33%** | 3.09% | 70.44%** | 2.04% | 72.36%* | 1.92% |
| W/o label consistency | 67.91%*** | 5.27% | 67.03%*** | 8.16% | 67.25%*** | 6.88% | 68.71%*** | 7.34% |
| W/o sample diversity | 65.35%*** | 9.40% | 66.90%*** | 8.37% | 65.84%*** | 9.17% | 68.26%*** | 8.04% |

Note: W/o: without. Imp: Improvement. Significance levels are denoted by * (0.05), ** (0.01) and *** (0.001) respectively.

### 5.3.3 Comparison of Performance in Different Classes

We further assessed HA-SOS's effectiveness in classifying various types of social support needs, as shown in Table 14. HA-SOS demonstrates significant improvements, especially in classifying emotional support and network support, outperforming baseline models by 20.36%–28.84% and 21.31%–33.49%, respectively. This is attributed to the smaller sample sizes for these categories, where HA-SOS's SSL and text data augmentation methods excel by effectively leveraging limited labeled data and generating high-quality synthetic samples. In contrast, for informational support, which already has a larger sample size, HA-SOS achieves a modest but consistent improvement. These results highlight HA-SOS's ability to address class imbalance and improve performance on underrepresented categories, making it particularly suitable for real-world applications where certain types may be less but equally important.

| Table 14. Comparison of Performance in Different Classes | | | | | | |
|---|---|---|---|---|---|---|
| Model | Informational support | | Emotional support | | Network support | |
| | AUC | Imp. | AUC | Imp. | AUC | Imp. |
| **HA-SOS (Ours)** | **60.67%** | - | **72.23%** | - | **66.72%** | - |
| RF (Mohasseb et al., 2018) | 54.57%*** | 11.18% | 56.06%*** | 28.84% | 49.98%*** | 33.49% |
| SVM (Zhang & Lee, 2003) | 54.61%*** | 11.10% | 59.76%*** | 20.87% | 52.35%*** | 27.45% |
| Tan et al. (2022) | 58.39%*** | 3.90% | 60.01%*** | 20.36% | 51.89%*** | 28.58% |
| Alasmari et al. (2023) | 60.00%* | 1.12% | 57.00%*** | 26.72% | 55.00%*** | 21.31% |
| GPT 4o + few shot | 59.11%*** | 2.64% | 59.32%*** | 21.76% | 50.88%*** | 31.13% |

Note: Imp: Improvement. Significance levels are denoted by * (0.05), ** (0.01) and *** (0.001) respectively.

### 5.3.4 Sensitivity Analysis

The performance of HA-SOS is influenced by several key hyperparameter. To understand this relationship, we conducted a sensitivity analysis focusing on the number of unlabeled samples and its effect on classification performance (see Table 15). The results reveal a clear trend: while increasing the volume of unlabeled data initially boosts performance, it eventually leads to diminished returns beyond an optimal threshold. Thus, an insufficient number of unlabeled samples restricts the model's learning capability, whereas excessive amounts may introduce noise. These findings underscore the importance of selecting an optimal volume of unlabeled data and provide practical guidance for configuring HA-SOS and similar SSL approaches by highlighting the need to balance data quantity and quality.

| Table 15. Effect of Number of Unlabeled Data on Performance | | | | |
|---|---|---|---|---|
| NO. unlabeled data | Precision | Recall | F1 | AUC |
| 5, 000 | 67.35% | 69.90% | 68.44% | 69.41% |
| **10, 000** | **71.49%** | **72.50%** | **71.88%** | **73.75%** |
| 20, 000 | 71.29% | 69.43% | 70.20% | 71.93% |

Another critical hyperparameter influencing performance is the number of answers included in the model. As shown in Table 16, increasing the number of answers is associated with a gradual decline in performance. This decline can be attributed to the higher likelihood of noise being introduced with additional answers. This insight underscores the importance of data quality over quantity in SSL frameworks like HA-SOS.

| Table 16.  Effect of Number of Answers on Performance | | | | |
|---|---|---|---|---|
| NO. of Answers | Precision | Recall | F1 | AUC |
| >=1 | **71.49%** | **72.50%** | **71.88%** | **73.75%** |
| >=2 | 69.48% | 70.76% | 70.01% | 71.18% |
| >=3 | 68.23% | 67.20% | 67.59% | 70.20% |
| >=4 | 68.96% | 69.53% | 69.16% | 70.38% |

To select generated samples that enhance model performance, we designed two evaluation metrics—label consistency and sample diversity—and a parameter, $\delta$, to balance their relative importance. Figure 5 illustrates the impact of varying $\delta$ on model performance. The optimal value of $\delta$ is found to be 0.4, indicating that the most useful generated samples are selected through a balanced combination of label consistency and sample diversity. These findings highlight the importance of jointly considering label consistency and sample diversity when evaluating LLM-generated data. This insight can guide future research in leveraging LLMs for data augmentation, emphasizing the need for robust evaluation criteria to ensure data quality.

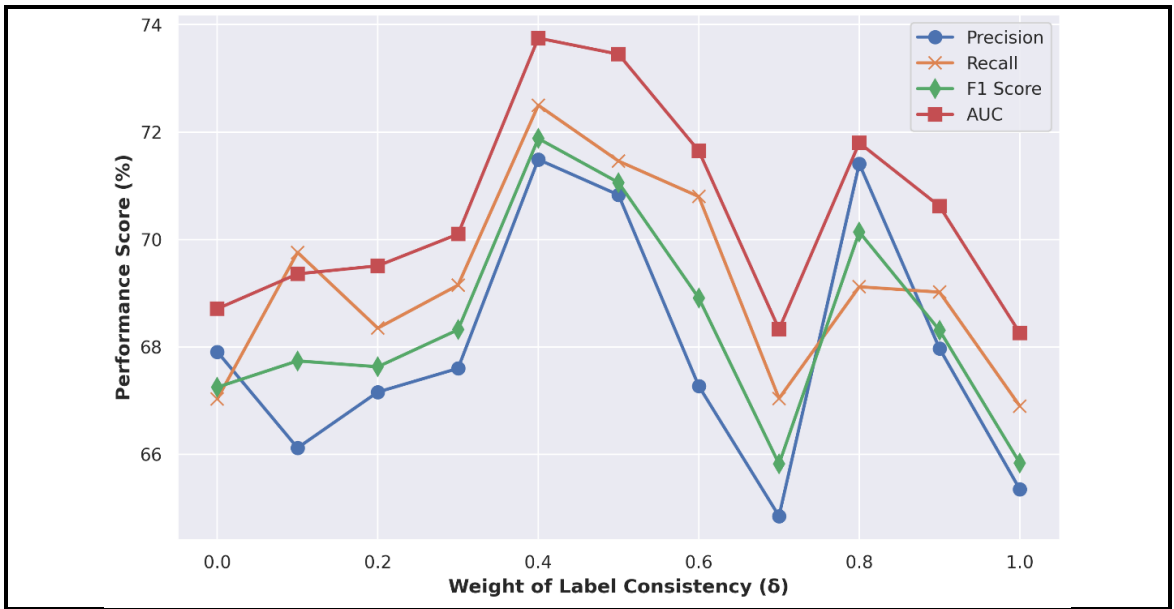

**Figure 4. Effect of Weight of Label Consistency and Sample Diversity on Performance**

To ensure the quality of LLM-generated samples, we assigned each sample a score based on label consistency and sample diversity and only included samples with scores above a predefined threshold $\eta$. Figure 6 summarizes the impact of varying $\eta$ on model performance. The results demonstrate a clear trend: as $\eta$ increases, the number of selected samples decreases, and model performance initially improves before eventually stabilizing. For example, at $\eta$ is 0.2, the model achieves its peak performance, as the selected samples strike an optimal balance between quantity and quality. However, when $\eta$ is too low, the inclusion of low-quality samples introduces noise, degrading performance. Conversely, when $\eta$ is too high, the reduced number of samples limits the model's ability to learn effectively. These findings highlight the importance of carefully selecting $\eta$ to control both the quantity and quality of samples generated.

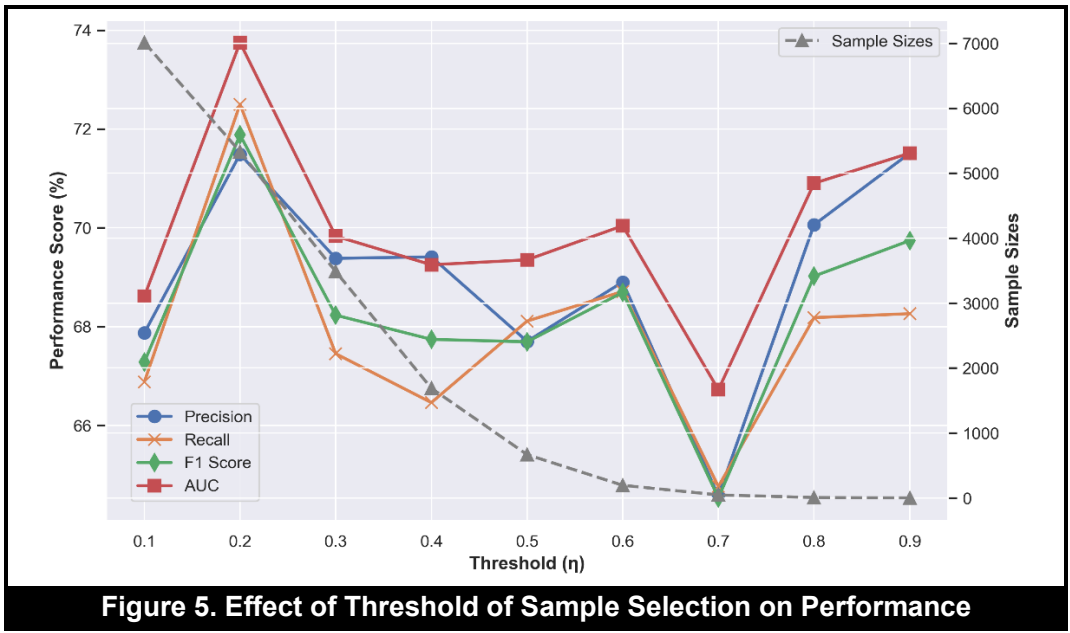

**Figure 5. Effect of Threshold of Sample Selection on Performance**

### 5.3.5 Interpretability Analysis

The dynamic 2D interaction kernel within HA-SOS plays a pivotal role in capturing interactions between question-and-answer texts. Table 17 presents several randomly selected question-answer pairs, highlighting potential interaction relationships identified by the kernel. For instance, the second row detects the supportive tone in the answer (e.g., "Bless your heart," "prayer"), aligning with the user's need for empathy and connection, characteristic of network support.

| Table 17. Typical Question-Answer Pairs | | |
|---|---|---|
| **Question** | **Answer** | **Type** |
| Living at home with family and this place is the Home from Hell most days. | Yes, stress can cause all the above symptoms and sounds like you have your share. | Emotional Support |
| Guess I just need to vent. Thanks for listening. | Bless your heart, I'll say a prayer for you. | Network Support |
| What can I do, in the meantime? Please help! | Do some research online for that and see if that doesn't match your symptoms. | Informational Support |
| Anyone know any more information on this than I do??? | Sickle cell disease is much more common in certain ethnic groups, but not unheard of in all. | Informational Support |

We also evaluated the interpretability of the quality-aware attention layer in HA-SOS. This layer is designed to assign more reasonable weights to answer texts by incorporating a constraint on the weight of the best answer. Table 18 compares the weights assigned by the traditional attention mechanism and our proposed quality-aware attention mechanism.

| Table 18. Case Studies on Quality-aware Attention Design | | | | |
|---|---|---|---|---|
| **Question** | **Answer** | **Is the Best Answer?** | **Traditional Attention** | **Our Attention** |
| I've been really sick....and whenever that happens....im so depressed about it...my boyfriend is coming over....and I wanted to look so nice for him...does anyone know any methods to make the sore go down in swelling :( … | I feel sorry for you. …. The cold sore virus is from the same family of viruses that causes genital herpes. …. It comes out in times of stress or when your immune system is a bit weak, ....  I hope this information is of some help to you. | Yes | 0.2331 | 0.2344 |
| | I know they are awful!  The best thing over-the-counter you can get for a coldsore is 'Abreva'.... it's pretty pricy (about 13 bucks for a small tube at Wal-Mart), but it's worth it... it really does help them heal quicker.... | No | 0.2445 | 0.2325 |
| | Try eating yogurt everyday or take L-Lysine vitamin.  It works for me - it is worth a try.  If you do start to get one and feel the tingle, put some ice on it and it should keep it small and then go away faster. | No | 0.2003 | 0.1062 |
| | no salty foods this is painfull but if you take menthol rubbing alcahol and litly dab it it will sooth th pain and reduce redness | No | 0.1786 | 0.2301 |
| | Try using some Carmex. | No | 0.1435 | 0.1968 |

For the question, the traditional attention mechanism assigns the highest weight (0.2445) to an answer (*"I know they are awful! The best thing over the counter..."*), while the best answer ("I feel sorry for you. I do know how you feel...") receives a lower weight (0.2331). In contrast, our quality-aware attention correctly assigns the highest weight (0.2344) to the best answer and adjusts the weights of other answers more appropriately. For instance, the answer *"Try eating yogurt everyday or take L-Lysine vitamin..."* receives a significantly reduced weight (0.1062 vs. 0.2003).

### 5.3.6 Generalizability Analysis

We extended our evaluation to a broader task: emotion classification (e.g., sadness, joy, fear) using the CHQ-SocioEmo dataset. Table 19 compares HA-SOS against a range of baseline methods and demonstrates that HA-SOS outperforms the best-performing baseline by 18.15% in Recall and 7.01% in AUC. This underscores HA-SOS's strong generalizability, suggesting its potential for broader applications in text classification tasks.

| Table 19. Comparison against Benchmark Methods for Emotion Classification | | | | | | | | |
|---|---|---|---|---|---|---|---|---|
| Model | Precision | Imp. | Recall | Imp. | F1 | Imp. | AUC | Imp. |
| DT (Bilal et al., 2016) | 37.12*** | 48.87% | 38.19*** | 35.51% | 37.61*** | 43.02% | 57.36*** | 19.96% |
| KNN (Zhang & Lee, 2003) | 42.54*** | 29.90% | 29.98*** | 72.62% | 35.11*** | 53.20% | 57.64*** | 19.38% |
| LR (Li et al., 2023) | 40.47*** | 36.55% | 44.65*** | 15.90% | 42.37*** | 26.95% | 60.35*** | 14.02% |
| MLP (Hamza et al., 2021) | 47.15*** | 17.20% | 42.88*** | 20.69% | 44.81*** | 20.04% | 62.74*** | 9.67% |
| NB (Momtazi, 2018) | 34.59*** | 59.76% | 42.93*** | 20.55% | 38.22*** | 40.74% | 56.77*** | 21.21% |
| RF (Mohasseb et al., 2018) | 53.18** | 3.91% | 15.77*** | 228.15% | 24.25*** | 121.81% | 55.4*** | 24.21% |
| SVM (Zhang & Lee, 2003) | 57.73 | -4.28% | 25.8*** | 100.58% | 35.53*** | 51.39% | 59.42*** | 15.80% |
| LSTM (Faris et al., 2022) | 44.74*** | 23.51% | 45.85*** | 12.87% | 44.96*** | 19.64% | 62.87*** | 9.45% |
| Bi-LSTM (Anhar et al., 2019) | 44.25*** | 24.88% | 44.70*** | 15.77% | 44.18*** | 21.75% | 62.29*** | 10.46% |
| CNN (Liu et al., 2019) | 52.6*** | 5.06% | 39.5*** | 31.01% | 44.93*** | 19.72% | 63.47*** | 8.41% |
| LSTM+ATT (Xia et al., 2018) | 46.96*** | 17.67% | 42.22*** | 22.57% | 44.2*** | 21.70% | 62.57*** | 9.97% |
| HAN (Yang et al., 2016) | 46.39*** | 19.12% | 44.19*** | 17.11% | 45.04*** | 19.43% | 63.03*** | 9.17% |
| Tan et al. (2022) | 47.02*** | 17.52% | 43.87*** | 17.96% | 45.1*** | 19.27% | 63.13*** | 9.00% |
| Liang et al. (2021) | 43.36*** | 27.44% | 43.10*** | 20.69% | 42.98*** | 25.15% | 61.59*** | 11.72% |
| Alasmari et al. (2023) | 55.64 | -0.68% | 33.6*** | 54.02% | 41.23*** | 30.46% | 61.8*** | 11.34% |
| **HA-SOS (Ours)** | **55.26** | - | **51.75** | - | **53.79** | - | **68.81** | - |

Note: Imp: Improvement. Significance levels are denoted by * (0.05), ** (0.01) and *** (0.001) respectively.

## 6  Discussion and Contributions

Patients increasingly turn to online health Q&A communities for diverse social support, yet the support received often misaligns with their specific needs. This highlights the necessity for a model capable of identifying social support needs in questions. To tackle the challenges of data scarcity and class imbalance in model training, we employed the computational design science paradigm to develop HA-SOS, a novel framework for social support needs classification. HA-SOS integrates an answer-enhanced semi-supervised learning approach with a text data augmentation technique that leverages LLMs and incorporates a reliability- and diversity-aware sample selection method. We rigorously evaluated HA-SOS against benchmark ML, DL, and LLMs methods for social support needs classification, underscoring its superior performance. Ablation studies were employed to validate the effectiveness of each innovative component. Additionally, we demonstrated HA-SOS's interpretability in analyzing typical question-answer pairs and critical answers, as well as its generalizability for emotion classification tasks. Our research contributes to the IS knowledge base and offers managerial and practical implications.

## 6.1 Contributions to the IS Knowledge Base

Novel IT artifacts often contribute prescriptive knowledge to the IS knowledge, offering guidance for future research (Abbasi et al., 2024; Hevner et al., 2004; Rai, 2017). Such contributions often include domain-specific implementations of IT artifacts and/or design principles with broader applicability. Our proposed HA-SOS is situated within the growing body of IS social support research. While existing literature has predominantly utilized empirical models to investigate the antecedents and outcomes of social support-related behaviors, there has been limited exploration of predictive analytics for social support. Considering these issues, this work aims to contribute a novel social support needs classification approach to the IS knowledge base. HA-SOS also follows four key design principles that are applicable beyond the social support needs classification: (1) combining complementary data sources (e.g., question and answer texts) during prediction tasks can enhance the quality of generated outputs (e.g., pseudo-labels); (2) enabling interaction between non-adjacent positions in multi-textual data (e.g., question-and-answer pairs) can capture more comprehensive patterns; (3) incorporating real user feedback (e.g., best answer selection) into the learning process can ensure the model prioritizes practical relevance and aligns with user preferences; (4) developing evaluation mechanisms (e.g., neighbor-based sample quality assessment) to filter out unreliable or redundant samples can improve the quality of retained data. These design principles could help guide the design of IT artifacts for E-commerce, health, E-education, and social media. Table 20 summarizes the framework components, the general design principles, the relevant IS literature to which each principle could contribute, and potential avenues for future research inquiry. Following this, we elaborate on how these design principles can offer value to each respective area of IS literature.

| Table 20 Design Principles for Selected Bodies and Classes of IS Research Inquiry | | | |
|---|---|---|---|
| **Framework component** | **General design principle** | **Relevant IS literature** | **Potential class of research inquiry** |
| Answer-enhanced semi-supervised learning | Combining complementary data sources during prediction tasks can enhance the quality of generated outputs | Social media | Using comments to enhance post classification |
| | | E-commerce | Using review to enhance product classification |
| Dynamic 2D kernel | Enabling interaction between non-adjacent positions in multi-textual data can capture more comprehensive patterns | Healthcare | Analyzing patterns between disparate sections of patient records, improving diagnostic tools |
| Quality-aware attention | Incorporating real user feedback into the learning process can ensure the model prioritizes practical relevance and aligns with user preferences | E-learning | Assigning higher weights to resources rated as "most helpful" by students to prioritize relevant content |
| Reliability- and diversity-aware sample selection | Developing evaluation mechanisms to filter out unreliable or redundant samples can improve the quality of retained data | E-commerce | Retain high-quality and diverse item listings to enhance relevance and user satisfaction |

### 6.1.1 Social Media

Social media platforms are used by billions of people worldwide, with users actively sharing posts on a wide range of topics. Accurately classifying these posts is critical for supporting users, managing online communities, and addressing societal challenges, as demonstrated by tasks such as fake news detection (Wei et al., 2022), suicide thought detection (Zhang et al., 2024b), and depression emotion detection (Chau et al., 2020; Peng et al., 2024; Zhang et al., 2024c). However, these tasks are hindered by the scarcity of labeled data due to the lack of clear labeling standards and the high cost of manual annotation. Scholars could consider Design Principle 1 to address these challenges by leveraging comments as complementary data sources. For instance, comments that fact-check a post can enhance fake news detection, while supportive comments can provide context for identifying emotional states in suicide or depression detection.

### 6.1.2 E-commerce

E-commerce platforms host millions of products, making accurate product classification essential for optimizing search results, and enhancing recommendation systems (He et al., 2019; Zhao et al., 2023). However, classifying products is challenging due to their increasing multifunctionality (e.g., a smartwatch serving as both a fitness tracker and a communication device) and the ambiguity of categorizing new or niche items. These challenges often result in misclassification, negatively impacting downstream tasks such as search and recommendation. Design Principle 1 addresses these issues by leveraging review data as a complementary information source (Chen et al., 2024). Reviews describe specific product features and user experiences, providing valuable context to clarify ambiguous product descriptions.

The product lists in search and recommendation systems play a critical role in shaping user satisfaction and engagement. Diversity in these lists is as important as relevance, as it exposes users to a broader range of products and enhances discovery (Shi et al., 2025; Yin et al., 2023). However, existing approaches often focus only on the relevance and diversity of items within the recommendation list, neglecting to compare them to the customer's purchase history (Zhou et al., 2023b). Design Principle 4 addresses this gap by evaluating the similarity between recommended items and the customer's purchase history to ensure reliability, while also measuring the diversity of recommendations relative to past purchases to avoid redundancy. For example, if a user frequently purchases running shoes, the system can prioritize recommending complementary products, such as sports socks or fitness trackers, rather than suggesting similar running shoes.

### 6.1.3 Healthcare

Predicting the risk of chronic diseases, such as diabetes or cardiovascular conditions, using clinical notes and patient records is essential for early intervention (Ben-Assuli & Padman, 2020; Lin et al., 2017; Yu et al., 2022). However, the fragmentation of patient data across different sections of medical records, such as symptoms, medical history, and diagnostic reports, poses significant challenges for accurate risk prediction. By applying Design Principle 2, interactions between non-adjacent data points in patient records can be analyzed to capture comprehensive patterns. For example, a system that identifies connections between a patient's past medical history and current symptoms can provide more accurate risk assessments. This approach enhances the accuracy of healthcare diagnostics, supports personalized treatment plans, ultimately leading to better patient outcomes and more efficient healthcare delivery.

### 6.1.4 E-learning

The rapid growth of e-learning has spurred significant interest in leveraging technology to enhance the structure and delivery of educational activities (Bauman & Tuzhilin, 2018; Li et al., 2024; Wambsganss et al., 2024). However, a key challenge lies in providing personalized learning resources—such as tailored video lectures, interactive exercises, or reading materials—that align with individual student needs and preferences. By incorporating Design Principle 3, a system that assigns higher weights to resources rated as "most helpful" by students can refine its attention mechanisms, enabling it to capture individual preferences. This approach ensures learners receive highly relevant materials, such as adaptive quizzes, which are continuously improved based on real user feedback.

**6.2    Managerial and Practical Implications**

Social support has gained significant attention in the IS field, particularly as online platforms increasingly become essential spaces for health-related communication and assistance (Liu et al., 2020b; Zhou et al., 2023a). The HA-SOS framework, designed to identify diverse social support needs in questions, enhances the quality of support provided to users, improves community engagement, and fosters better health outcomes. This framework offers substantial value to key stakeholders, including community askers, answerers, healthcare professionals, and platform managers. Below, we elaborate on the practical implications of HA-SOS for each stakeholder.

**Community askers** rely on health Q&A platforms to seek timely and relevant answers to their health-related questions. A primary challenge they face is articulating their needs clearly, as poorly framed questions often lead to mismatched answers. HA-SOS addresses these challenges by classifying questions into specific support categories, guiding askers to better understand and express their needs when formulating questions. This ensures that their questions are more likely to receive appropriate and helpful responses, enhancing their overall experience and satisfaction.

**Community answerers** aim to efficiently identify and respond to questions that align with their expertise and interests. However, they often struggle with information overload, making it difficult to locate suitable questions. HA-SOS mitigates this challenge by categorizing questions into specific support needs, enabling answerers to quickly identify relevant questions. For instance, an expert could easily locate and respond to questions requiring emotional support, while another expert could focus on providing detailed informational resources.

**Healthcare professionals** depend on timely and accurate insights into patients' needs to enhance treatment outcomes. Traditional methods, such as surveys or interviews, often face resistance from patients due to privacy concerns (Xie et al., 2022). In contrast, online health Q&A communities encourage users to disclose their needs. However, the vast volume of information generated on these platforms makes manual analysis impractical. The HA-SOS framework addresses this challenge by automatically classifying questions into specific social support needs and extract key sentences or words related to each need.

**Platform managers** are responsible for fostering a supportive community by effectively connecting askers and answerers, ensuring user satisfaction, and maintaining high engagement levels. HA-SOS supports these goals by automating the classification of questions into specific social support needs and routing them to the most appropriate answerers. Additionally, the accurate classification of social support needs enables multiple downstream tasks, such as expert recommendation (matching questions to qualified answerers) and question recommendation (suggesting relevant questions to answerers) (Yuan et al., 2020).

## 7 Conclusion and Future Directions

Social support has a profound impact on health, yet the support patients receive online is not always beneficial and can be harmful. The effectiveness of such support largely depends on its alignment with the asker's specific needs. Classifying social support needs is therefore essential for ensuring individuals receive timely and appropriate assistance. However, existing challenges, such as data scarcity and class imbalance in model training, have limited the performance of existing machine learning models in identifying social support needs in questions.

In this study, we adopted the computational design science paradigm to develop a novel framework, the Hybrid Approach for SOcial Support need classification (HA-SOS). HA-SOS integrates an answer-enhanced semi-supervised learning approach, a text data augmentation technique leveraging with reliability- and diversity-aware sample selection mechanism. Extensive empirical evaluations on real-world datasets demonstrate that HA-SOS significantly outperforms state-of-the-art question classification methods, alternative semi-supervised learning approaches, and text data augmentation techniques. Ablation studies further underscore the critical contributions of the proposed components, such as the dynamic 2D interaction kernel and quality-aware attention layer. In practice, our HA-SOS framework facilitates online Q&A platform managers and users to better understand users' social support needs, enabling them to provide timely, personalized answers and interventions.

Despite its strengths, our study has several limitations that warrant future research. First, HA-SOS is currently designed for single-turn Q&A scenarios and does not account for multi-turn dialogues. Future work could incorporate dialogue modeling techniques, enabling it to better handle evolving conversations and capture the dynamic nature of multi-turn exchanges. Second, HA-SOS currently prioritizes only the "best answer," ignoring other potentially valuable answers that may not receive the highest ratings but still offer useful insights. This limitation could be overcome by considering other user feedback such as number of likes, replies, and retweets. Third, HA-SOS is primarily suited for structured Q&A datasets, while many online platforms follow a post-comment structure where the question is often implicit. Future work could explore these less structured formats to detect social support needs within posts and comments.

**REFERENCES**

Abbasi, A., Parsons, J., Pant, G., Liu Sheng, O. R., & Sarker, S. (2024). Pathways for Design Research on Artificial Intelligence. *Information Systems Research*, *35*(2), 441–459. https://doi.org/10.1287/isre.2024.editorial.v35.n2

Alasmari, A., Kudryashov, L., Yadav, S., Lee, H., & Demner-Fushman, D. (2023). CHQ- SocioEmo: Identifying Social and Emotional Support Needs in Consumer-Health Questions. *Scientific Data*,

*10*(1), 1–12. https://doi.org/10.1038/s41597-023-02203-1

Ampel, B. M., Samtani, S., Zhu, H., & Chen, H. (2024). Creating Proactive Cyber Threat Intelligence With Hacker Exploit Labels: A Deep Transfer Learning Approach. *MIS Quarterly*, *48*(1), 137–166. https://doi.org/10.25300/MISQ/2023/17316

Anhar, R., Adji, T. B., & Akhmad Setiawan, N. (2019). Question classification on question-answer system using bidirectional-LSTM. *Proceedings - 2019 5th International Conference on Science and Technology, ICST 2019*. https://doi.org/10.1109/ICST47872.2019.9166190

Baird, A., Angst, C., & Oborn, E. (2020). MIS Quarterly Research Curation on Health Information Technology. *MIS Quarterly*, *2007*(May 2018), 1–17.

Bauman, K., & Tuzhilin, A. (2018). Recommending remedial learning materials to students by filling their knowledge gaps. *MIS Quarterly*, *42*(1), 313–332. https://doi.org/10.25300/MISQ/2018/13770

Bayer, M., Kaufhold, M. A., & Reuter, C. (2022). A Survey on Data Augmentation for Text Classification. *ACM Computing Surveys*, *55*(7), 1–44. https://doi.org/10.1145/3544558

Ben-Assuli, O., & Padman, R. (2020). Trajectories of repeated readmissions of chronic disease patients: Risk stratification, profiling, and prediction. *MIS Quarterly*, *44*(1), 201–226. https://doi.org/10.25300/MISQ/2020/15101

Bilal, M., Israr, H., Shahid, M., & Khan, A. (2016). Sentiment classification of Roman-Urdu opinions using Naïve Bayesian, Decision Tree and KNN classification techniques. *Journal of King Saud University - Computer and Information Sciences*, *28*(3), 330–344. https://doi.org/10.1016/j.jksuci.2015.11.003

Brock, R. L., & Lawrence, E. (2009). Too Much of a Good Thing: Underprovision Versus Overprovision of Partner Support. *Journal of Family Psychology*, *23*(2), 181–192. https://doi.org/10.1037/a0015402

Chang, Y., Wang, X., Wang, J., Wu, Y., Yang, L., Zhu, K., Chen, H., Yi, X., Wang, C., Wang, Y., Ye, W., Zhang, Y., Chang, Y., Yu, P. S., Yang, Q., & Xie, X. (2024). A Survey on Evaluation of Large Language Models. *ACM Transactions on Intelligent Systems and Technology*, *15*(3). https://doi.org/10.1145/3641289

Chau, M., Li, T. M. H., Wong, P. W. C., Xu, J. J., Yip, P. S. F., & Chen, H. (2020). Finding people with emotional distress in online social media: A design combining machine learning and rule-BASED classification. *MIS Quarterly*, *44*(2), 933–956. https://doi.org/10.25300/MISQ/2020/14110

Chavoshinejad, J., Seyedi, S. A., Akhlaghian Tab, F., & Salahian, N. (2023). Self-supervised semi-supervised nonnegative matrix factorization for data clustering. *Pattern Recognition*, *137*, 109282. https://doi.org/10.1016/j.patcog.2022.109282

Chen, G., Huang, L., Xiao, S., Zhang, C., & Zhao, H. (2024). Attending to Customer Attention: A Novel Deep Learning Method for Leveraging Multimodal Online Reviews to Enhance Sales Prediction. *Information Systems Research*, *35*(2), 829–849. https://doi.org/10.1287/isre.2021.0292

Chen, L., Baird, A., & Straub, D. (2019). Fostering Participant Health Knowledge and Attitudes: An Econometric Study of a Chronic Disease-Focused Online Health Community. *Journal of Management Information Systems*, *36*(1), 194–229. https://doi.org/10.1080/07421222.2018.1550547

Chen, L., Baird, A., & Straub, D. (2020). A linguistic signaling model of social support exchange in online health communities. *Decision Support Systems*, *130*, 113233. https://doi.org/10.1016/j.dss.2019.113233

Chen, L., Zhang, D., & Levene, M. (2012). Understanding user intent in Community Question Answering. *WWW'12 - Proceedings of the 21st Annual Conference on World Wide Web Companion*, 823–828. https://doi.org/10.1145/2187980.2188206

Chintagunta, B., Katariya, N., Amatriain, X., & Kannan, A. (2021). Medically Aware GPT-3 as a Data Generator for Medical Dialogue Summarization. *Proceedings of Machine Learning Research*, *149*(1), 354–372. https://doi.org/10.18653/v1/2021.nlpmc-1.9

Chiu, C. M., Huang, H. Y., Cheng, H. L., & Sun, P. C. (2015). Understanding online community citizenship behaviors through social support and social identity. *International Journal of Information Management*, *35*(4), 504–519. https://doi.org/10.1016/j.ijinfomgt.2015.04.009

Cui, H., Wang, G., Li, Y., & Welsch, R. E. (2022). Self-training method based on GCN for semi-supervised short text classification. *Information Sciences*, *611*, 18–29. https://doi.org/10.1016/j.ins.2022.07.186

Dai, C., Wu, J., Monaghan, J. J. M., Li, G., Peng, H., Becker, S. I., & Mcalpine, D. (2023). Semi-Supervised EEG Clustering With Multiple Constraints. *IEEE Transactions on Knowledge and Data Engineering*, *35*(8), 8529–8544. https://doi.org/10.1109/TKDE.2022.3206330

Devlin, J., Chang, M. W., Lee, K., & Toutanova, K. (2019). BERT: Pre-training of deep bidirectional transformers for language understanding. *NAACL HLT 2019 - 2019 Conference of the North American Chapter of the Association for Computational Linguistics: Human Language Technologies - Proceedings of the Conference*, *1*, 4171–4186.

Dong, X., & de Melo, G. (2019). A robust self-learning framework for cross-lingual text classification. *EMNLP-IJCNLP 2019 - 2019 Conference on Empirical Methods in Natural Language Processing and 9th International Joint Conference on Natural Language Processing, Proceedings of the Conference*, 6306–6310. https://doi.org/10.18653/v1/d19-1658

Duarte, J. M., & Berton, L. (2023). A review of semi-supervised learning for text classification. In *Artificial Intelligence Review* (Vol. 56, Issue 9). Springer Netherlands. https://doi.org/10.1007/s10462-023-10393-8

Ebrahimi, M., Nunamaker, J. F., & Chen, H. (2020). Semi-Supervised Cyber Threat Identification in Dark Net Markets: A Transductive and Deep Learning Approach. *Journal of Management Information Systems*, *37*(3), 694–722. https://doi.org/10.1080/07421222.2020.1790186

Emadi, M., Tanha, J., Shiri, M. E., & Aghdam, M. H. (2021). A Selection Metric for semi-supervised learning based on neighborhood construction. *Information Processing and Management*, *58*(2), 102444. https://doi.org/10.1016/j.ipm.2020.102444

Faris, H., Habib, M., Faris, M., Alomari, A., Castillo, P. A., & Alomari, M. (2022). Classification of Arabic healthcare questions based on word embeddings learned from massive consultations: a deep learning approach. *Journal of Ambient Intelligence and Humanized Computing*, *13*(4), 1811–1827. https://doi.org/10.1007/s12652-021-02948-w

Gao, F., Zhu, J., Wu, L., Xia, Y., Qin, T., Cheng, X., Zhou, W., & Liu, T. Y. (2019). Soft contextual data augmentation for neural machine translation. *57th Annual Meeting of the Association for Computational Linguistics, Proceedings of the Conference*, 5539–5544. https://doi.org/10.18653/v1/p19-1555

Goh, J. M., Gao, G., & Agarwal, R. (2016). The creation of social value: can an online health community reduce rural-urban health disparities? *MIS Quarterly*, *40*(1), 247–263. https://doi.org/10.25300/MISQ/2016/40.1.11

Gray, C. E., Spector, P. E., Lacey, K. N., Young, B. G., Jacobsen, S. T., & Taylor, M. R. (2020). Helping may be Harming: unintended negative consequences of providing social support. *Work & Stress*, *34*(4), 359–385. https://doi.org/10.1080/02678373.2019.1695294

Gregor, S., & Hevner, A. R. (2013). Positioning and presenting design science research for maximum impact. *MIS Quarterly*, *37*(2), 337–355. https://doi.org/10.25300/MISQ/2013/37.2.01

Guo, L. Z., & Li, Y. F. (2022). Class-Imbalanced Semi-Supervised Learning with Adaptive Thresholding. *Proceedings of Machine Learning Research*, *162*, 8082–8094.

Guo, T., Bardhan, I. R., Ding, Y., Zhang, S., Guo, T., Bardhan, I. R., & Ding, Y. (2024). An Explainable Artificial Intelligence Approach Using Graph Learning to Predict Intensive Care Unit Length of Stay. *Information Systems Research*. https://doi.org/10.1287/isre.2023.0029

Hajmohammadi, M. S., Ibrahim, R., & Selamat, A. (2014). Bi-view semi-supervised active learning for cross-lingual sentiment classification. *Information Processing and Management*, *50*(5), 718–732. https://doi.org/10.1016/j.ipm.2014.03.005

Hamza, A., En-Nahnahi, N., Zidani, K. A., & El Alaoui Ouatik, S. (2021). An arabic question classification method based on new taxonomy and continuous distributed representation of words. *Journal of King Saud University - Computer and Information Sciences*, *33*(2), 218–224. https://doi.org/10.1016/j.jksuci.2019.01.001

He, J., Fang, X., Liu, H., & Li, X. (2019). Mobile App Recommendation: An Involvement-Enhanced Approach. *MIS Quarterly*, *43*(3), 827–850. https://doi.org/10.25300/MISQ/2019/15049

Hevner, A. R., March, S. T., Park, J., & Ram, S. (2004). Design science in information systems research.

*MIS Quarterly*, *28*(1), 75–105. https://doi.org/10.2307/25148625

Hu, C., Liu, B., Ye, Y., & Li, X. (2023). Fine-grained classification of drug trafficking based on Instagram hashtags. *Decision Support Systems*, *165*, 113896. https://doi.org/10.1016/j.dss.2022.113896

Huang, K. Y., Chengalur-Smith, I. S., & Pinsonneault, A. (2019). Sharing is caring: Social support provision and companionship activities in healthcare virtual support communities. *MIS Quarterly*, *43*(2), 395–423. https://doi.org/10.25300/MISQ/2019/13225

Kang, L., Liu, J., Liu, L., Zhou, Z., & Ye, D. (2021). Semi-supervised emotion recognition in textual conversation via a context-augmented auxiliary training task. *Information Processing and Management*, *58*(6), 102717. https://doi.org/10.1016/j.ipm.2021.102717

Kim, Y. (2014). Convolutional Neural Networks for Sentence Classification. *Proceedings of the 2014 Conference on Empirical Methods in Natural Language Processing (EMNLP)*, 1746–1751. https://doi.org/10.3115/v1/D14-1181

Kowshik, S. S., Divekar, A., & Malik, V. (2024). CorrSynth - A Correlated Sampling Method for Diverse Dataset Generation from LLMs. *Proceedings of the 2024 Conference on Empirical Methods in Natural Language Processing (EMNLP)*, 16076–16095.

Kundu, D., Pal, R. K., & Mandal, D. P. (2020). Preference enhanced hybrid expertise retrieval system in community question answering services. *Decision Support Systems*, *129*, 113164. https://doi.org/10.1016/j.dss.2019.113164

Lee, J., Rajtmajer, S., Srivatsavaya, E., & Wilson, S. (2023). Online Self-Disclosure, Social Support, and User Engagement During the COVID-19 Pandemic. *ACM Transactions on Social Computing*, *6*(3–4), 1–31. https://doi.org/10.1145/3617654

Leimeister, J. M., Schweizer, K., Leimeister, S., & Krcmar, H. (2008). Do virtual communities matter for the social support of patients? Antecedents and effects of virtual relationships in online communities. *Information Technology and People*, *21*(4), 350–374. https://doi.org/10.1108/09593840810919671

Lekwijit, S., Terwiesch, C., Asch, D. A., & Volpp, K. G. (2024). Evaluating the Efficacy of Connected Healthcare: An Empirical Examination of Patient Engagement Approaches and Their Impact on Readmission. *Management Science*, *70*(6), 3417–3446. https://doi.org/10.1287/mnsc.2023.4865

Li, A. T., Liu, D., Xu, S. X., & Yi, C. (2024). Interleaved Design for E-learning: Theory, Design, and Empirical Findings. *MIS Quarterly*, *48*(4), 1363–1394. https://doi.org/10.25300/misq/2023/17206

Li, C., Li, X., & Ouyang, J. (2021). Semi-supervised text classification with balanced deep representation distributions. *ACL-IJCNLP 2021 - 59th Annual Meeting of the Association for Computational Linguistics and the 11th International Joint Conference on Natural Language Processing, Proceedings of the Conference*, 5044–5053. https://doi.org/10.18653/v1/2021.acl-long.391

Li, Q., Zhao, S., Zhao, S., & Wen, J. (2023). Logistic Regression Matching Pursuit algorithm for text classification. *Knowledge-Based Systems*, *277*, 110761. https://doi.org/10.1016/j.knosys.2023.110761

Li, Y., Su, L., Chen, J., & Yuan, L. (2017). Semi-supervised learning for question classification in CQA. *Natural Computing*, *16*(4), 567–577. https://doi.org/10.1007/s11047-016-9554-5

Liang, Y., Li, H., Guo, B., Yu, Z., Zheng, X., Samtani, S., & Zeng, D. D. (2021). Fusion of heterogeneous attention mechanisms in multi-view convolutional neural network for text classification. *Information Sciences*, *548*, 295–312. https://doi.org/10.1016/j.ins.2020.10.021

Ligthart, A., Catal, C., & Tekinerdogan, B. (2021). Analyzing the effectiveness of semi-supervised learning approaches for opinion spam classification. *Applied Soft Computing*, *101*, 107023. https://doi.org/10.1016/j.asoc.2020.107023

Lin, M.-P., Wu, J. Y.-W., You, J., Chang, K.-M., Hu, W.-H., & Xu, S. (2018). Association between online and offline social support and internet addiction in a representative sample of senior high school students in Taiwan: The mediating role of self-esteem. *Computers in Human Behavior*, *84*, 1–7. https://doi.org/10.1016/j.chb.2018.02.007

Lin, Y.-K., Chen, H., Brown, R. A., Li, S.-H., & Yang, H.-J. (2017). Healthcare predictive analytics for risk profiling in chronic care: a bayesian multitask learning approach. *MIS Quarterly*, *41*(2), 473–495.

Lin, Y.-K., & Fang, X. (2021). First, Do No Harm: Predictive Analytics to Reduce In-Hospital Adverse

Events. *Journal of Management Information Systems*, *38*, 1122–1149. https://doi.org/10.2139/ssrn.3273203

Liu, J., Yang, Y., Lv, S., Wang, J., & Chen, H. (2019). Attention-based BiGRU-CNN for Chinese question classification. *Journal of Ambient Intelligence and Humanized Computing*, *0123456789*. https://doi.org/10.1007/s12652-019-01344-9

Liu, X., Alan Wang, G., Fan, W., & Zhang, Z. (2020). Finding useful solutions in online knowledge communities: A theory-driven design and multilevel analysis. *Information Systems Research*, *31*(3), 731–752. https://doi.org/10.1287/ISRE.2019.0911

Liu, X., Zhang, B., Susarla, A., & Padman, R. (2020). Go to youtube and call me in the morning: Use of social media for chronic conditions. *MIS Quarterly*, *44*(1), 257–283. https://doi.org/10.25300/MISQ/2020/15107

Madabushi, H. T., & Lee, M. (2016). High accuracy rule-based question classification using question syntax and semantics. *Proceedings of COLING 2016, the 26th International Conference on Computational Linguistics: Technical Papers*, 1220–1230.

Mallikarjuna, C., & Sivanesan, S. (2022). Question classification using limited labelled data. *Information Processing and Management*, *59*(6), 103094. https://doi.org/10.1016/j.ipm.2022.103094

Martin, M. (2022). *Facebook stats that matter to marketers in 2022*.

Meng, Y., Shen, J., Zhang, C., & Han, J. (2018). Weakly-supervised neural text classification. *International Conference on Information and Knowledge Management, Proceedings*, 983–992. https://doi.org/10.1145/3269206.3271737

Meng, Y., Shen, J., Zhang, C., & Han, J. (2019). Weakly-supervised hierarchical text classification. *33rd AAAI Conference on Artificial Intelligence, AAAI 2019, 31st Innovative Applications of Artificial Intelligence Conference, IAAI 2019 and the 9th AAAI Symposium on Educational Advances in Artificial Intelligence, EAAI 2019*, 6826–6833. https://doi.org/10.1609/aaai.v33i01.33016826

Merdan, S., Barnett, C. L., Denton, B. T., Montie, J. E., & Miller, D. C. (2021). OR practice-data analytics for optimal detection of metastatic prostate cancer. *Operations Research*, *69*(3), 774–794. https://doi.org/10.1287/OPRE.2020.2020

Mi, C., Xie, L., & Zhang, Y. (2022). Improving data augmentation for low resource speech-to-text translation with diverse paraphrasing. *Neural Networks*, *148*, 194–205. https://doi.org/10.1016/j.neunet.2022.01.016

Mirzaei, M. S., Meshgi, K., & Sekine, S. (2023). What is the Real Intention behind this Question? Dataset Collection and Intention Classification. *Proceedings of the Annual Meeting of the Association for Computational Linguistics*, *1*, 13606–13622. https://doi.org/10.18653/v1/2023.acl-long.761

Mohasseb, A., Bader-El-Den, M., & Cocea, M. (2018). Question categorization and classification using grammar based approach. *Information Processing and Management*, *54*(6), 1228–1243. https://doi.org/10.1016/j.ipm.2018.05.001

Momtazi, S. (2018). Unsupervised Latent Dirichlet Allocation for supervised question classification. *Information Processing and Management*, *54*(3), 380–393. https://doi.org/10.1016/j.ipm.2018.01.001

Nambisan, P. (2011). Information seeking and social support in online health communities: Impact on patients' perceived empathy. *Journal of the American Medical Informatics Association*, *18*(3), 298–304. https://doi.org/10.1136/amiajnl-2010-000058

Oh, H. J., Ozkaya, E., & Larose, R. (2014). How does online social networking enhance life satisfaction? the relationships among online supportive interaction, affect, perceived social support, sense of community, and life satisfaction. *Computers in Human Behavior*, *30*, 69–78. https://doi.org/10.1016/j.chb.2013.07.053

Oliveira, W. D. G. de, & Berton, L. (2023). A systematic review for class-imbalance in semi-supervised learning. *Artificial Intelligence Review*, *56*, 2349–2382. https://doi.org/10.1007/s10462-023-10579-0

Padmanabhan, B., Fang, X., Sahoo, N., & Burton-Jones, A. (2022). Machine Learning in Information Systems Research. *MIS Quarterly*, *46*(1), iii–xix.

Park, Y., & Ho, J. C. (2021). Tackling Overfitting in Boosting for Noisy Healthcare Data. *IEEE Transactions on Knowledge and Data Engineering*, *33*(7), 2995–3006.

https://doi.org/10.1109/TKDE.2019.2959988

Pavlinek, M., & Podgorelec, V. (2017). Text classification method based on self-training and LDA topic models. *Expert Systems with Applications*, *80*, 83–93. https://doi.org/10.1016/j.eswa.2017.03.020

Pellicer, L. F. A. O., Ferreira, T. M., & Costa, A. H. R. (2023). Data augmentation techniques in natural language processing. *Applied Soft Computing*, *132*, 109803. https://doi.org/10.1016/j.asoc.2022.109803

Peng, F., Zhang, D., & Yan, Z. (2024). Digital Phenotyping-based Depression Detection in the Presence of Comorbidity: An Uncertainty Reasoning Approach. *Journal of Management Information Systems*, *41*(4), 931–957. https://doi.org/10.1080/07421222.2024.2415770

Rai, A. (2017). Editor's Comments: Diversity of Design Science Research. *MIS Quarterly*, *41*(1), iii–xviii.

Rana, J., Mukku, S., Yenigalla, P., Soni, M., Aggarwal, C., & Patange, R. (2023). Weakly supervised hierarchical multi-task classifcation of customer questions. *Proceedings of the Annual Meeting of the Association for Computational Linguistics*, *5*, 786–793. https://doi.org/10.18653/v1/2023.acl-industry.75

Ray, S. K., Singh, S., & Joshi, B. P. (2010). A semantic approach for question classification using WordNet and Wikipedia. *Pattern Recognition Letters*, *31*(13), 1935–1943. https://doi.org/10.1016/j.patrec.2010.06.012

Sahu, G., Rodriguez, P., Laradji, I. H., Atighehchian, P., Vazquez, D., & Bahdanau, D. (2022). Data Augmentation for Intent Classification with Off-the-shelf Large Language Models. *Proceedings of the Annual Meeting of the Association for Computational Linguistics*, 47–57. https://doi.org/10.18653/v1/2022.nlp4convai-1.5

Shandwick, W. (2018). *The Great American Search for Healthcare Information*. https://cms.webershandwick.com/wp-content/uploads/2023/01/Healthcare-Infor-Search-Report.pdf

Shi, L., Liu, J., Li, Y., & Foutz, N. Z. (2025). Ephemeral State-Dependent Recommendation for Digital Content. *Information Systems Research*, 1–14. https://doi.org/10.1287/isre.2022.664

Shin, H., Hou, T., Park, K., Park, C. K., & Choi, S. (2013). Prediction of movement direction in crude oil prices based on semi-supervised learning. *Decision Support Systems*, *55*(1), 348–358. https://doi.org/10.1016/j.dss.2012.11.009

Silva, J., Coheur, L., Mendes, A. C., & Wichert, A. (2011). From symbolic to sub-symbolic information in question classification. *Artificial Intelligence Review*, *35*(2), 137–154. https://doi.org/10.1007/s10462-010-9188-4

Simester, D., Timoshenko, A., & Zoumpoulis, S. I. (2020). Targeting prospective customers: Robustness of machine-learning methods to typical data challenges. *Management Science*, *66*(6), 2495–2522. https://doi.org/10.1287/mnsc.2019.3308

Sohn, K., Berthelot, D., Li, C. L., Zhang, Z., Carlini, N., Cubuk, E. D., Kurakin, A., Zhang, H., & Raffel, C. (2020). FixMatch: Simplifying semi-supervised learning with consistency and confidence. *Advances in Neural Information Processing Systems*, 596–608.

Tan, Z., Chen, J., Kang, Q., Zhou, M., Abusorrah, A., & Sedraoui, K. (2022). Dynamic Embedding Projection-Gated Convolutional Neural Networks for Text Classification. *IEEE Transactions on Neural Networks and Learning Systems*, *33*(3), 973–982. https://doi.org/10.1109/TNNLS.2020.3036192

Tarekegn, A. N., Giacobini, M., & Michalak, K. (2021). A review of methods for imbalanced multi-label classification. *Pattern Recognition*, *118*, 107965. https://doi.org/10.1016/j.patcog.2021.107965

Thirunavukarasu, A. J., Ting, D. S. J., Elangovan, K., Gutierrez, L., Tan, T. F., & Ting, D. S. W. (2023). Large language models in medicine. *Nature Medicine*, *29*(8), 1930–1940. https://doi.org/10.1038/s41591-023-02448-8

Trepte, S., Dienlin, T., & Reinecke, L. (2015). Influence of Social Support Received in Online and Offline Contexts on Satisfaction With Social Support and Satisfaction With Life: A Longitudinal Study. *Media Psychology*, *18*(1), 74–105. https://doi.org/10.1080/15213269.2013.838904

Vaux, A. (1988). *Social support: Theory, research, and intervention*. Praeger publishers.

Verma, D. A., Joshi, R. S., Joshi, S. A., & Susladkar, O. K. (2021). Study of Similarity Measures as Features

in Classification for Answer Sentence Selection Task in Hindi Question Answering: Language-Specific v/s Other Measures. *Proceedings of the 35th Pacific Asia Conference on Language, Information and Computation, PACLIC 2021*, 715–724.

Wambsganss, T., Janson, A., Söllner, M., Koedinger, K., & Leimeister, J. M. (2024). Improving Students' Argumentation Skills Using Dynamic Machine-Learning–Based Modeling. In *Information Systems Research*. http://pubsonline.informs.org/journal/isre. https://doi.org/10.1287/isre.2021.0615

Wang, G., Chen, G., Zhao, H., Zhang, F., Yang, S., & Lu, T. (2021). Leveraging Multisource Heterogeneous Data for Financial Risk Prediction: A Novel Hybrid-Strategy-Based Self-Adaptive Method. *MIS Quarterly*, *45*(4), 1949–1998. https://doi.org/10.25300/MISQ/2021/16118

Wang, G., Cheng, Y., Xia, Y., Ling, Q., & Wang, X. (2023). A Bayesian Semisupervised Approach to Keyword Extraction with Only Positive and Unlabeled Data. *INFORMS Journal on Computing*, *35*(3), 675–691. https://doi.org/10.1287/ijoc.2023.1283

Wang, X., Zhao, K., Cha, S., Amato, M. S., Cohn, A. M., Pearson, J. L., Papandonatos, G. D., & Graham, A. L. (2019). Mining user-generated content in an online smoking cessation community to identify smoking status: A machine learning approach. *Decision Support Systems*, *116*, 26–34. https://doi.org/10.1016/j.dss.2018.10.005

Wang, Y., Wang, J., Yao, T., Li, M., & Wang, X. (2020). How does social support promote consumers' engagement in the social commerce community? The mediating effect of consumer involvement. *Information Processing and Management*, *57*(5), 102272. https://doi.org/10.1016/j.ipm.2020.102272

Wei, X., Zhang, Z., Zhang, M., Chen, W., & Zeng, D. D. (2022). Combining Crowd and Machine Intelligence To Detect False News on Social Media1. *MIS Quarterly*, *46*(2), 977–1008. https://doi.org/10.25300/MISQ/2022/16526

Whitehouse, C., Choudhury, M., & Aji, A. F. (2023). LLM-powered Data Augmentation for Enhanced Crosslingual Performance. *EMNLP 2023 - 2023 Conference on Empirical Methods in Natural Language Processing, Proceedings*, 671–686. https://doi.org/10.18653/v1/2023.emnlp-main.44

Xia, W., Zhu, W., Liao, B., Chen, M., Cai, L., & Huang, L. (2018). Novel architecture for long short-term memory used in question classification. *Neurocomputing*, *299*, 20–31. https://doi.org/10.1016/j.neucom.2018.03.020

Xie, J., Liu, X., Zeng, D. D., & Fang, X. (2022). Understanding Medication Nonadherence from Social Media: A Sentiment-Enriched Deep Learning Approach. *MIS Quarterly*, *46*(1), 341–372. https://doi.org/10.25300/MISQ/2022/15336

Xu, J. J., Chen, D., Chau, M., Li, L., & Zheng, H. (2022). Peer-to-Peer Loan Fraud Detection: Constructing Features from Transaction Data. *MIS Quarterly*, *46*(3), 1777–1792. https://doi.org/10.25300/MISQ/2022/16103

Yan, L. (2020). The Kindness of Commenters: An Empirical Study of the Effectiveness of Perceived and Received Support for Weight-Loss Outcomes. *Production and Operations Management*, *29*(6), 1448–1466. https://doi.org/10.1111/poms.13171

Yan, L. L. (2018). Good Intentions, Bad Outcomes: The Effects of Mismatches between Social Support and Health Outcomes in an Online Weight Loss Community. *Production and Operations Management*, *27*(1), 9–27. https://doi.org/10.1111/poms.12793

Yan, L., & Tan, Y. (2014). Feeling blue? Go online: An empirical study of social support among patients. *Information Systems Research*, *25*(4), 690–709. https://doi.org/10.1287/isre.2014.0538

Yang, X., Li, G., & Huang, S. S. (2017). Perceived online community support, member relations, and commitment: Differences between posters and lurkers. *Information and Management*, *54*(2), 154–165. https://doi.org/10.1016/j.im.2016.05.003

Yang, X., Song, Z., King, I., & Xu, Z. (2023). A Survey on Deep Semi-Supervised Learning. *IEEE Transactions on Knowledge and Data Engineering*, *35*(9), 8934–8954. https://doi.org/10.1109/TKDE.2022.3220219

Yang, Y., Wang, Y., Wang, L., & Meng, J. (2022). PostCom2DR: Utilizing information from post and comments to detect rumors. *Expert Systems with Applications*, *189*, 116071. https://doi.org/10.1016/j.eswa.2021.116071

Yang, Z., Yang, D., Dyer, C., He, X., Smola, A., & Hovy, E. (2016). Hierarchical Attention Networks for Document Classification. *Proceedings of the 2016 Conference of the North American Chapter of the Association for Computational Linguistics: Human Language Technologies*, 1480–1489.

Yilmaz, S., & Toklu, S. (2020). A deep learning analysis on question classification task using Word2vec representations. *Neural Computing and Applications*, *32*(7), 2909–2928. https://doi.org/10.1007/s00521-020-04725-w

Yin, K., Fang, X., Chen, B., & Sheng, O. R. L. (2023). Diversity Preference-Aware Link Recommendation for Online Social Networks. *Information Systems Research*, *34*(4), 1398–1414. https://doi.org/10.1287/isre.2022.1174

Yoo, K. M., Park, D., Kang, J., Lee, S. W., & Park, W. (2021). GPT3Mix: Leveraging Large-scale Language Models for Text Augmentation. *Findings of the Association for Computational Linguistics: EMNLP 2021*, 2225–2239. https://doi.org/10.18653/v1/2021.findings-emnlp.192

You, Y., Tsai, C. H., Li, Y., Ma, F., Heron, C., & Gui, X. (2023). Beyond Self-diagnosis: How a Chatbot-based Symptom Checker Should Respond. *ACM Transactions on Computer-Human Interaction*, *30*(4), 1–44. https://doi.org/10.1145/3589959

Yu, S., Chai, Y., Chen, H., Sherman, S. J., & Brown, R. A. (2022). Wearable Sensor-Based Chronic Condition Severity Assessment: an Adversarial Attention-Based Deep Multisource Multitask Learning Approach. *MIS Quarterly*, *46*(3), 1355–1394. https://doi.org/10.25300/MISQ/2022/15763

Yu, S., Chai, Y., Samtani, S., Liu, H., & Chen, H. (2024). Motion Sensor–Based Fall Prevention for Senior Care: A Hidden Markov Model with Generative Adversarial Network Approach. *Information Systems Research*, *35*(1), 1–15. https://doi.org/10.1287/isre.2023.1203

Yuan, S., Zhang, Y., Tang, J., Hall, W., & Cabotà, J. B. (2020). Expert finding in community question answering: a review. *Artificial Intelligence Review*, *53*(2), 843–874. https://doi.org/10.1007/s10462-018-09680-6

Zepeda, E. D., & Sinha, K. K. (2016). Toward an effective design of behavioral health care delivery: An empirical analysis of care for depression. *Production and Operations Management*, *25*(5), 952–967. https://doi.org/10.1111/poms.12529

Zhang, D., & Lee, W. S. (2003). Question Classification using Support Vector Machines. *Proceedings of the 26th Annual International ACM SIGIR Conference on Research and Development in Information Retrieval*, 26–32. https://doi.org/10.1145/860435.860443

Zhang, D., Zhang, K., Yang, Y., & Schweidel, D. A. (2024). TM-OKC: An Unsupervised Topic Model for Text in Online Knowledge Communities. *MIS Quarterly*, *48*(3), 931–978. https://doi.org/10.25300/MISQ/2023/17885

Zhang, D., Zhou, L., Tao, J., Zhu, T., & Gao, G. (Gordon). (2024). KETCH: A Knowledge-Enhanced Transformer-Based Approach to Suicidal Ideation Detection from Social Media Content. *Information Systems Research*, *August*. https://doi.org/10.1287/isre.2021.0619

Zhang, W., Xie, J., Zhang, Z., & Liu, X. (2024). Depression Detection Using Digital Traces on Social Media: A Knowledge-aware Deep Learning Approach. *Journal of Management Information Systems*, *41*(2), 546–580.

Zhao, X., Fang, X., He, J., & Huang, L. (2023). Exploiting Expert Knowledge for Assigning Firms To Industries: a Novel Deep Learning Method. *MIS Quarterly*, *47*(3), 1147–1176. https://doi.org/10.25300/MISQ/2022/17171

Zhou, J., Kishore, R., Amo, L., & Ye, C. (2022). Description and Demonstration Signals As Complements and Substitutes in an Online Market for Mental Health Care. *MIS Quarterly*, *46*(4), 2055–2084. https://doi.org/10.25300/MISQ/2022/16122

Zhou, J., Zhang, Q., Zhou, S., Li, X., & Zhang, X. (2023). Unintended Emotional Effects of Online Health Communities: a Text Mining-Supported Empirical Study. *MIS Quarterly*, *47*(1), 195–226. https://doi.org/10.25300/MISQ/2022/17018

Zhou, T., Wang, Y., Yan, L., & Tan, Y. (2023). Spoiled for Choice? Personalized Recommendation for Healthcare Decisions: A Multiarmed Bandit Approach. *Information Systems Research*, *34*(4), 1493–1512. https://doi.org/10.1287/isre.2022.1191

Zhu, H., Samtani, S., Brown, R. A., & Chen, H. (2021). A deep learning approach for recognizing activity of daily living (adl) for senior care: Exploiting interaction dependency and temporal patterns. *MIS Quarterly*, *45*(2), 859–896. https://doi.org/10.25300/MISQ/2021/15574

# APPENDIX A

**LLMs Prompt for Data Augmentation**

**CHQ_prompt**

### Context

You are a data generator capable of producing new samples and corresponding labels based on a few given examples, i.e., Few-shot Samples. Your data generation focuses on mental health-related questions and answers. Each item in the list below contains a question from an online mental health community and the corresponding best answer. The labels are "informational_support_need", "emotional_support_need", and "social_support_need". Each item indicates whether the question reflects a need for one or more types of support.

### Instruction

Please generate 20 new samples, each consisting of a question and corresponding labels. These samples should meet the following requirements:

The total number of True instances for each label should be as equal as possible across the entire set of samples. All three labels ("informational_support_need", "emotional_support_need", "social_support_need") should not be True simultaneously in any instance.

Ensure the novelty of the generated samples; they should not just rephrase existing ones but instead offer new perspectives or scenarios.

The generated samples must be of high quality, reflecting realistic and relatable situations in online mental health discussions.

### Few-shot Samples

{}

20 new samples:

""

## APPENDIX B

**Benchmark Model Specifications**

For each machine learning method, we used grid-search in Scikit-Learn[6] to search for the best parameters for classification task. The key parameter settings are summarized in Table B1.

| Table B1. Key Hyperparameter Settings Involved in the Benchmark Model | | |
|---|---|---|
| **Hyperparameter** | | **Value** |
| Decision Tree | Criterion | Gini |
| | Max_depth | 8 |
| K-Nearest Neighbor | N_neighbors | 7 |
| | Weights | distance |
| Logistic Regression | Penalty | L2 |
| Multi-layer Perceptron | Hidden_layer_sizes | 256 |
| | Max_iter | 200 |
| | Learning_rate | 0.01 |
| | Batch_size | 64 |
| | Solver | Adam |
| Naïve Bayes | Additive (Laplace/Lidstone) smoothing parameter | 1 |
| Random Forest | N_estimators | 100 |
| | Criterion | Gini |
| | Max_depth | 5 |
| Support Vector Machine | Regularization parameter | 1.0 |
| | Kernel | Radial basis function |
| | Degree | 3 |
| | Gamma | Scale |

For each deep learning method, we used PyTorch to construct and train our models. Following Ampel et al. (2024), we adjusted the parameters based on practices in related literature and each model maintained the same layers, including embedding, convolutional, batch normalization, and dropout, to ensure comparability. The dropout rate was set consistently at 0.4 across all models. The kernel size of convolutional layer was set to 3 and a ReLU activation function was employed. For all RNN-based networks, the number of recurrent layers was 1.

---

[6] https://scikit-learn.org/stable/index.html